\documentclass{aa}  

\usepackage{natbib, multirow}
\usepackage{graphicx}
\usepackage{txfonts}
\usepackage{color}
\usepackage{pdflscape}
\definecolor{aipred}{rgb}{0.63,0.0,0.16}
\definecolor{aipdarkblue}{rgb}{0.0,0.25,0.47}
\usepackage[colorlinks=true,urlcolor=blue,citecolor=aipdarkblue]{hyperref}

\newcommand{\oiii}{[\ion{O}{iii}]$\lambda5007$}

\newcommand{\hii}{\ion{H}{ii}}
\begin{document}

   \title{Towards Precision Cosmology With Improved PNLF Distances Using VLT-MUSE}

   \subtitle{III. Impact of Stellar Populations in Early-Type Galaxy}

   \author{Azlizan A. Soemitro \inst{1,2}
          \and
          Lucas M. Valenzuela\inst{3}
          \and
          Martin M. Roth \inst{1,2}
          \and
          Robin Ciardullo  \inst{4,5}
          \and
          George H. Jacoby \inst{6}
          \and
          Magda Arnaboldi \inst{7}
          \and
          Guilherme S. Couto\inst{1} 
          \and
          C. Jakob Walcher\inst{1}
          }

   \institute{Leibniz-Institut für Astrophysik Potsdam (AIP),
              An der Sternwarte 16, 14482 Potsdam, Germany\\
              \email{asoemitro@aip.de}
         \and
            Institut für Physik und Astronomie, Universität Potsdam, Karl-Liebknecht-Str. 24/25, 14476 Potsdam, Germany
         \and
            Universitäts-Sternwarte, Fakultät für Physik, Ludwig-Maximilians Universität München, Scheinerstr. 1, 81679 München, Germany
         \and
            Department of Astronomy \& Astrophysics, The Pennsylvania State University, University Park, PA 16802, USA
         \and
            Institute for Gravitation and the Cosmos, The Pennsylvania State University, University Park, PA 16802, USA
         \and
            NSF’s NOIRLab, 950 N. Cherry Ave., Tucson, AZ 85719, USA
         \and
            European Southern Observatory, Karl-Schwarzschild-Straße 2, 85748, Garching, Germany
             }

   \date{Received ; accepted }

  \abstract
   {}
   {Distance measurements using the planetary nebula luminosity function (PNLF) rely on the bright-end power-law cut-off magnitude ($M^*$), which is defined by a number of the \oiii-brightest planetary nebulae (PNe). In early-type galaxies (ETGs), the formation of these PNe is enigmatic; the population is typically too old to form the expected $M^*$ PNe from single star evolution. We aim to give a solution to this problem.}
   {We selected five ETGs with known MUSE-PNLF distances. The MUSE instrument allows us to calculate the PNLF and consistently investigate the underlying stellar populations. Using stellar population synthesis, we derive the population age, star formation history, metallicity, and alpha abundance. We compare these parameters to the PNLF variables: $M^*$ and luminosity-specific PN number at the top 0.5 mag of the PNLF ($\alpha_{0.5}$). We also compare our results with PNe In Cosmological Simulations (PICS) model applied to Magneticum Pathfinder analogue galaxies.}
   {The average mass-weighted ages and metallicities of our observations are typically old ($9 <\mathrm{Age}< 13.5$ Gyr) and rather metal-rich ($-0.4 <\mathrm{[M/H]}< +0.2$). We find $M^*$ to be independent of age and metallicity in these ages and metallicity intervals. We discover a positive correlation between $\alpha_{0.5}$ values and the mass fraction of stellar population ages of 2--10 Gyr, implying that most of the PNe originate from stars with intermediate ages. Similar trends are also found in the PICS analogue galaxies.}
   {We show that the presence of at least $\sim 2\%$ of stellar mass younger than 10 Gyr is, in principle, sufficient to form the $M^*$ PNe in ETGs. We also present observing requirements for an ideal PNLF distance determination in ETGs.}


\maketitle

%

\section{Introduction}

The Planetary Nebula Luminosity Function (PNLF) was proposed as a standard candle for extragalactic distance determinations more than 35 years ago \citep{1989ApJ...339...39J, 1989ApJ...339...53C}. It relies on the empirical fact that narrow-band photometry in the \oiii, of a sufficiently large sample of planetary nebulae (PNe) in any galaxy shows a luminosity function with a well defined ``knee'' and cut-off magnitude at the bright end ($M^*$) that can serve as a standard candle. As explained in a review by \citet{2022FrASS...9.6326C}, in the first two decades of its use, the technique produced almost a hundred papers on precision distance determinations out to 20 Mpc. However, interest in the method declined in the following years. In the early 2010s, direct calibration of supernova type Ia parent galaxies using Cepheids yielded $\sim 3$\% precision for the Hubble constant, $H_0$ \citep{2011ApJ...730..119R}. This superseded the need for intermediate candles, such as the PNLF\null. Subsequently, Cepheid measurements in recent years pushed the $H_0$ precision down to $\sim 1$\% \citep{2022ApJ...934L...7R}, which was crucial to the emergence of the Hubble Tension \citep{2021CQGra..38o3001D}. 

With the advent of MUSE \citep[Multi Unit Spectral Explorer;][]{2010SPIE.7735E..08B} at the ESO Very Large Telescope (VLT), a new parameter space for narrow band imaging became available.  MUSE is an integral field spectrograph with a 1~arcmin$^2$ field of view, excellent image quality, relatively high spectral resolution ($R \sim 2000$) and high efficiency.  The instrument offers many advantages over traditional narrow-band surveys for the detection of \oiii~point sources, including a much narrower effective bandpass and spectral information which immediately helps with the rejection of interlopers, such as \hii~regions, supernova remnants, and other bright emission-line sources \citep{2008ApJ...683..630H,2009ApJ...703..894H, 2017ApJ...834..174K, 2018MNRAS.476.1909A}.

Several authors (e.g., \citealp{2021A&A...653A.167S, 2022MNRAS.511.6087S, 2023A&A...672A.148C, 2023A&A...671A.142S}) have found the instrument to exceed the sensitivity and distance range accessible by classical narrow-band imaging. \citet{2018A&A...618A...3R} showed that, by placing simulated point sources inside real MUSE data cubes, \oiii~photometry can be accomplished to magnitudes as faint as $m_{5007} = 29$. More importantly, \citet{2021ApJ...916...21R}, henceforth Paper I, discovered that the galaxy background continuum light, which must be subtracted during PN aperture photometry, can be estimated with great precision using the full spectrum information present in the data cube.  Unlike the classical on-band/off-band measurements for the narrow-band filter technique, this fact enables the application of a self-calibrating background subtraction technique, called DELF (differential emission line filter). In addition to the benefits of MUSE mentioned above, DELF provides an additional factor of two improvement detection over other detection techniques, as demonstrated with three benchmark galaxies (Paper~I).

Encouraged by this finding and using the DELF technique, \citet[][hereafter Paper~II]{2024ApJS..271...40J} analysed MUSE data cubes for 20 galaxies that were retrieved from the ESO archive, and determined their respective PNLFs. Despite some data being taken in less-than-ideal observing conditions and the calibration issues associated with data taken for other purposes, this proof-of-principle study demonstrated that PNLF measurements with MUSE can yield 5\% precision distances to individual galaxies. It also shows that the PNLF method can be pushed to distances as far as $\sim 40$~Mpc with ground-based telescopes. This is comparable to the distance range of the best space-based observations of Cepheids and the Tip of the Red Giant Branch (TRGB), and well within the range of surface brightness fluctuations (SBF) distances.   Based on the studies presented in Paper I and II, and considering optimized observing strategies with MUSE, a renaissance of the PNLF is now possible.

Unlike Cepheid and TRGB distances that rely on well-established stellar physics \citep[][and references therein]{2024A&ARv..32....4B, 2018SSRv..214..113B}, there are still many uncertainties concerning the theoretical basis of the PNLF\null. The main aspect to understand is whether $M^*$ has some small dependence on stellar population that has not been measured.  While the effect of metallicity on PNLF distances is reasonably well understood \citep{1992ApJ...389...27D, 2002ApJ...577...31C}, its insensitivity to age is more problematic. This issue is critical, especially in old systems such as elliptical or lenticular galaxies. 

According to most initial-final mass (IFMR) relations \citep[e.g.,][]{2018ApJ...866...21C, 2024MNRAS.527.3602C}, the observed PNLF cut-off at $M^* = -4.54$ implies a population of progenitor stars that are $\sim 3$ Gyr or younger \citep{2018NatAs...2..580G, 2022FrASS...9.6326C}; such stars are exceedingly rare in Population~II systems. \citet{2005ApJ...629..499C} attempted to explain this phenomenon through binary star evolution, and more recently, \citet{2025A&A...699A.371V} argued that shifts in the IFMR with helium and metal abundance can achieve the canonical $M^*$ for populations as old as 10 Gyr. Additionally, \citet{2025ApJ...983..129J} suggested that a small amount of scatter in the IFMR, consistent with that used to explain the horizontal branch morphologies of old star clusters, could produce and maintain the cut-off PNe, even in the oldest systems. If true, this predicts a trend where the production rate of \oiii-bright, and therefore the luminosity specific number of such objects \citep[the $\alpha$-parameter,][]{2005ApJ...629..499C, 2006MNRAS.368..877B}, declines with increasing population age. Higher $\alpha$-parameters have been reported in the outer haloes of three old elliptical galaxies (M87, M49 and M105), and linked to bluer, lower metallicity environments \citep{2015A&A...579A.135L, 2017A&A...603A.104H, 2020A&A...642A..46H}.
These studies examine how the PNLF is affected by stellar population changes, traced through colour gradients and colour–magnitude diagrams of resolved stellar populations. 

In this third paper in the MUSE-PNLF series, we attempt to shed light on the physics behind the PNLF's bright-end cut-off in early-type galaxies (ETGs).  To do this, we use five ETGs with publicly available MUSE data from the ESO Archive \citep{2022SPIE12186E..0DR}.  For each galaxy, we use the MUSE data to measure the PNLF's cut-off magnitude $M^*$ and normalization parameter $\alpha$ (the luminosity specific PN number) as a function of radius, and correlate these parameters' values with the galaxies' stellar populations age and metallicity, as derived from the population synthesis analyses of the galaxies' MUSE spectra \citep{2023arXiv231109176E}. We also study the correlation between the PNLF and stellar population in the simulated galaxies of \citet{2025A&A...699A.371V} to test whether the trends seen in the observations and simulations are in agreements. Investigating how the stellar population relates to $M^*$ and the \oiii-bright PN production rate is an important next step to address the enigma of the empirically inferred invariance of the absolute magnitude $M^*$ of the PNLF bright cut-off.

The paper is organized as follows: Sect. \ref{sec:data} explains the selection criteria for the galaxy sample used in this study. The overall methodology is presented in Sect. \ref{sec:method}. The technical details on the PNLF fitting, the stellar population synthesis, and the simulation analogue selections are described in Sect. \ref{sec:PNLF_fits}, Sect. \ref{sec:SPS}, and Sect. \ref{sec:simanalogue}, respectively. The results are presented in Sect. \ref{sec:results}. Our discussion on the PNLF cut-off follows in Sect. \ref{sec:discussions}. We close the paper with conclusions and an outlook in Sect. \ref{sec:conclusion}. 


\section{Data and Sample Selection} \label{sec:data}

The MUSE data that we retrieved from the ESO Archive were observed using the Nominal Wide Field Mode without adaptive optics (WFM-NOAO-N\null). This mode covers a $1\arcmin \times 1\arcmin$ field-of-view with a sampling of 0$\farcs$2 per spaxel, and covers the wavelength range 4750~\AA\ $< \lambda < 9300$~\AA, with a spectral resolution that ranges from $R \sim 1770$ in the blue to $R \sim 3590$ in the far red. 

To have a high quality measurement of the PNLF cut-off, $\gtrsim 30$ PNe must be present in the brightest magnitude of the luminosity function (Paper I and II). If we are to split the PNe within a galaxy into different sub-populations, then this number must be increased, preferably by a factor of two.  This limits the number of MUSE archival systems available for analysis.

Our selection is based on the ETGs whose PNLF distances were derived in Papers I and II\null.  In these papers, five galaxies have enough bright PNe for our study:  NGC\,1052, NGC\,1351, NGC\,1380, NGC\,1399, and NGC\,1404\null. With more than 80 PNe per galaxy, we can define two PN sub-populations. In principle, NGC\,1399 has a sufficient PN number to be split into three or more sub-populations. However, we left it at two groups for uniformity. The overview of the MUSE data is given in Table \ref{tab:dataset overview}.

\renewcommand{\arraystretch}{1.3}
\begin{table*}
\caption{Dataset overview.}
\label{tab:dataset overview}
\centering
\begin{tabular}{l l c c c c}
\hline
Galaxy & ESO Archive ID [Program ID] & Exp. time [s] & Seeing & $N_{\mathrm{PNe}}$\tablefootmark{a} & $N_{\mathrm{PNe,0.5}}$\tablefootmark{b} \\
\hline
NGC 1052 & ADP.2019-10-05T19-19-48.724 (0103.B-0837) & 1685 & 0$\farcs$61 & 86 & 26 \\
NGC 1351 & ADP.2017-12-12T15-38-27.863 (296.B-5054) & 3379 & 0$\farcs$69 & 86 & 16 \\
NGC 1380 & ADP.2017-01-30T14-39-33.249 (296.B-5054) & 3369 & 0$\farcs$63 & 112 & 23 \\
NGC 1399 & ADP.2017-03-27T13-16-27.827 (094.B-0298; 094.B-0903) & 954 & 0$\farcs$81 & 232 & 37 \\
NGC 1404 & ADP.2017-12-13T01:47:07.213 (296.B-5054) & 3287 & 0$\farcs$70 & 107 & 21 \\
\hline
\end{tabular}
\tablefoot{
\tablefoottext{a}{NGC\,1380 from Paper I. The other galaxies from Paper II.}
\tablefoottext{b}{Number of PNe at the top 0.5 magnitude of the PNLF.}
}
\end{table*}

\begin{table*}
\caption{Ellipse definitions for the PN sub-population.}
\label{tab:my_label}
\centering
\begin{tabular}{c c c c c c c c c}
\hline
Galaxy & $D$ [Mpc]\tablefootmark{a} & $R_e$ [$''$] & $R_e$ [kpc] & PA [degree] & $\epsilon$ & $a_1$ [$''$] & $a_2$ [$''$] & $a_3$ [$''$] \\
\hline
NGC\,1052  & 17.9 & 21.9  & 1.9  & 112.89 & 0.29 & 9.87  & 28.41 & 59.93 \\
NGC\,1351  & 19.0 & 27.6  & 2.5  & 141.56 & 0.36 & 5.20  & 24.83 & 60.30 \\
NGC\,1380  & 16.6 & 72.7  & 5.9  & 186.17 & 0.27 & 8.53  & 52.86 & 222.51 \\
NGC\,1399  & 17.6 & 346   & 29.5 & 111.31 & 0.12 & 15.20 & 44.28 & 111.12 \\
NGC\,1404  & 18.8 & 201.5 & 18.4 & 159.09 & 0.16 & 5.61  & 25.16 & 57.72 \\
\hline
\end{tabular}
\tablefoot{
\tablefoottext{a}{PNLF distances from Paper I and II.}
}
\end{table*}

For NGC\,1052 and NGC\,1399, we used the same datasets as in Paper II\null. NGC\,1052 is a mosaic of two pointings from the same observing program, with a similar image quality of $\sim$ 0$\farcs$6 and the same exposure time of 2788 seconds. The MUSE data of NGC\,1399 is also a mosaic, but one constructed from five pointings with a range of seeings (0$\farcs$6 $-$ 1$\farcs$7) and exposure times (900 $-$ 1800 seconds) from two different programs. As discussed in Paper II, this mosaic was compromised by imperfect flat-field corrections and consequently, inconsistent noise levels. Following Paper II, we excluded the inner $15"$, due to incompleteness caused by high surface brightness. For NGC\,1351, NGC\,1380, and NGC\,1404, we only employed the single pointings close to the nucleus, where most of the PN candidates are located. We found that the nuclear pointings of these galaxies are sufficient to comply with our selection criteria.


\section{Methodology} \label{sec:method}

\begin{figure*}[h!]
    \centering
    \includegraphics[width=0.75\linewidth]{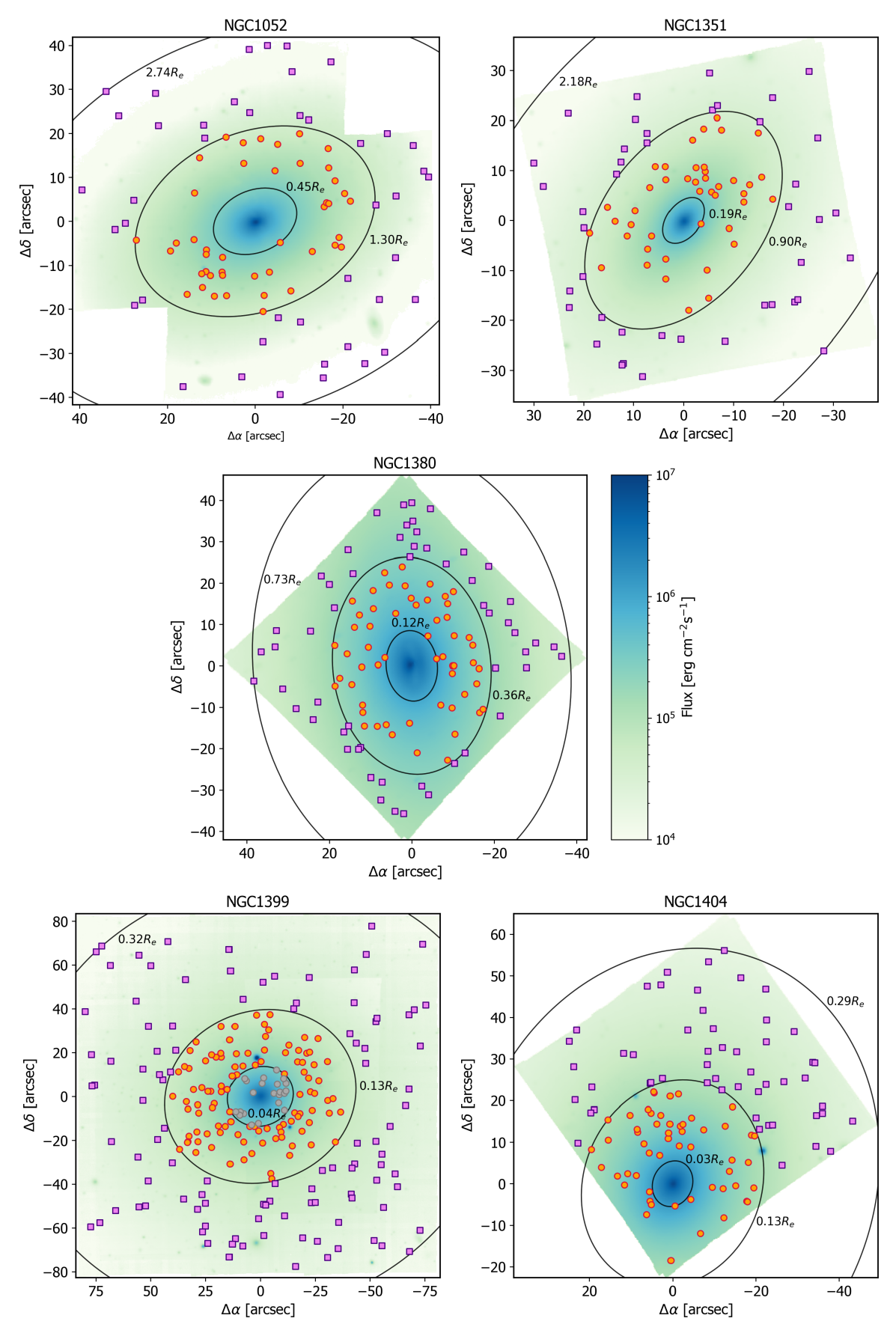}
    \caption{PN sub-population definitions for our galaxies. The orange circles represent the positions of the inner region PNe, and the purple square points denote the positions of the outer sub-population. The galaxy surface brightness is logarithmically scaled in a same manner for all galaxies. Following Paper II, we exclude the inner $15\arcsec$ of NGC\,1399, as the PNe sample in that region is incomplete.}
    \label{fig:ellipse}
\end{figure*}

The definition of the two PN sub-population is based on the PNe's isophotal galactocentric distance. First, we created synthetic $R$-band images from the MUSE datacube using the Python package \textit{synphot} \citep{pey_lian_lim_2020_3971036}. Using the images, we used \textit{photutils.isophote} \citep{larry_bradley_2024_12585239} to define three ellipses with position angle PA and ellipticity $\epsilon$:  an inner ellipse with semi-major axis $a_1$, within which PN detections are compromised by the galaxy's high surface brightness, a middle ellipse with semi-major axis $a_2$, which encloses half the observed PN population, and an outer ellipse $a_3$, defined by the PN furthest from the nucleus. To convert the semi-major axes into effective radius units, we adopted $R_e$ from \citet{2017MNRAS.464.4611F} for NGC\,1052 and \citet{2019A&A...627A.136I} for the rest. These parameters are listed in Table \ref{tab:my_label} and illustrated in Fig. \ref{fig:ellipse}. The position angles and ellipticities are in good agreement with those found by previous surface photometry studies \citep{1987AJ.....94.1508J,2019A&A...623A...1I}.

Using these definitions, we fit the PNLF of each region to obtain the cut-off $M^*$ and the $\alpha$-parameter, which we explained in Sect. \ref{sec:PNLF_fits}. We then investigate the characteristics of the underlying stellar population for the defined regions in terms of age, $\mathrm{[M/H]}$ and $[\alpha/\mathrm{Fe}]$, using stellar population synthesis. The technical details are described in Sect. \ref{sec:SPS}. While the relation between the underlying stellar population and the PNLF cut-off has previously been studied using selected line-indices \citep{2005ApJ...629..499C}, our analysis is much more comprehensive, as it not only yields metallicity information, but also the fraction of stellar mass in three age groups ($t < 2$ Gyr, $2 \leq t \leq 10$ Gyr, and $t > 10$ Gyr). Our goal is to determine the most likely population progenitors of the PNe near the PNLF cut-off.

Finally, following our population synthesis analysis, we use the simulations of \citet{2025A&A...699A.371V} to compare their theoretical predictions with our observational results. The aim is to have a self-consistent study between the underlying stellar population and the PNLF cut-off, from both observational and theoretical perspectives. Details on the selection criteria for the cosmological simulations are explained in Sect. \ref{sec:simanalogue}.

\section{Fitting the PNLF} \label{sec:PNLF_fits}

We parametrized the shape of the cut-off to the PNLF using the analytic expression first given by \citet{1989ApJ...339...53C}; although more generalised expressions of the PNLF are available \citep[e.g.,][]{2013A&A...558A..42L} this simple form is sufficient for our purposes. Briefly stated, we fit the PN magnitudes to
\begin{equation}
\phi_1(M) \propto e^{0.307 M} \{ 1 - e^{3 (M^* - M)} \}
\label{eq:pnlf}
\end{equation}
via the methodology described by \citet{2023ApJ...950...59C}.   These fits assume the \citet{2011ApJ...737..103S} value for foreground Galactic extinction and use the published estimates of the galaxies' surface brightness and position-dependent line-of-sight velocity dispersions to take into account the possibility of PN superpositions in the high surface brightness regions of the galaxies.  This results in a probability distribution function of two variables: the galaxy's true distance modulus $(m-M)_0$ \citep[assuming the PNLF's bright-end cut-off is at $M^* = -4.54$;][] {2022FrASS...9.6326C} and the luminosity function's normalization, $\alpha_{2.5}$. Note that to compute the values of the latter variable, we use an assumed $V$-band bolometric correction of $BC_V = -0.85$ \citep{2006MNRAS.368..877B}; the $\alpha_{2.5}$ value is computed as the number of PNe within 2.5~mag of $M^*$ per unit bolometric luminosity of the sampled stellar population. However, for the relevance of distance determination, we are only interested in the PNe in the top 0.5~mag of the luminosity function, which define the PNLF cut-off. We therefore scale our observed $\alpha_{2.5}$ values to $\alpha_{0.5}$ following the recipe of \citet{2005ApJ...629..499C} for our analysis.

For both the distance modulus and PN luminosity density, our errors account for the statistical errors of the fits only.  Systematic uncertainties, such as those associated with the foreground reddening estimate and the photometric zero point of MUSE are not included.  However, since our analysis is differential, comparing two regions within the same galaxy, the omission of these terms will not change our results. 

\section{Stellar Population Synthesis} \label{sec:SPS}
\subsection{Population Models} \label{subsec:popmodels}

We utilized the semi-empirical Medium-resolution Isaac Newton Telescope Library of Empirical Spectra \citep[sMILES;]{2021MNRAS.504.2286K} simple stellar population (SSP) models \citep{2023MNRAS.523.3450K} to analyze the galactic spectra contained on the MUSE frames. The sMILES library uses as its basis the MILES empirical spectra \citep{2006MNRAS.371..703S, 2011A&A...532A..95F}, but includes fully theoretical corrections which allow for variations in the abundance of $\alpha$-process elements, with 
$[\alpha/\mathrm{Fe}]$ families of $-0.2$, 0.0, +0.2, +0.4, and +0.6 \citep{2021MNRAS.504.2286K}.  The library itself covers the wavelength range $\lambda\lambda 3540.5 - 7409.6$~{\AA}\ at a spectral resolution of 2.51~{\AA\null}.   To calculate single stellar populations (SSP), \citet{2023MNRAS.523.3450K} adopted solar scaled isochrones \citep{2004ApJ...612..168P} for $[\alpha/\mathrm{Fe}]$ = $-0.2$, 0.0, +0.2 and $\alpha$-enhanced isochrones \citep{2006ApJ...642..797P} for $[\alpha/\mathrm{Fe}]$ = +0.4, +0.6. The models were computed in 10 metallicity steps $[\mathrm{M/H}]$ = $-1.79$, $-1.49$, $-1.26$, $-0.96$, $-0.66$, $-0.35$, $-0.25$, +0.06, +0.15, +0.26, as defined by \citet{grevesse1993cosmic}, and 53 age steps between 0.03 and 14 Gyr. The SSPs were provided with five variations of the initial mass function (IMF), but only the \citet{2001MNRAS.322..231K} IMF is used in our analysis.

As a consistency check, we also applied two other SSP models to our data. First, we tested the models from \citet{2015MNRAS.449.1177V}, that are based on the original MILES spectral library. These SSPs include $[\alpha/\mathrm{Fe}]$ = 0.0, 0.4 models calculated using a correction factor obtained from a fully theoretical stellar library \citep{2005A&A...443..735C, 2007MNRAS.382..498C}.  Note that in these models the abundance correction was applied to the SSP model spectra as a whole, in contrast to \citet{2023MNRAS.523.3450K} where their corrections were applied to the individual star spectra before constructing the sMILES SSPs. Second, we also tested the semi-empirical model of \citet{2009MNRAS.398L..44W}, which used the theoretical library from \citet{2007MNRAS.382..498C} to correct their SSP fluxes as a function of $[\mathrm{Fe/H}]$ and $[\alpha/\mathrm{Fe}]$. Their base SSP models are adopted from \citet{2003MNRAS.344.1000B}, \citet{2003A&A...402..433L}, and \citet{1999ApJ...513..224V} and include  ages between 2$-$13 Gyr, abundances of $[\mathrm{Fe/H}]$ =  $-$0.5, $-$0.25, 0.0, 0.2, and $[\alpha/\mathrm{Fe}]$ = 0.0, 0.2, 0.4. The model resolution is a constant FWHM = 1.0~{\AA\null}. Detailed discussion on the comparison between different SSP models are presented in Appendix \ref{appendix:SSP_comparison}. 

\subsection{SSP fitting}
\label{subsec:fitting}

Our observational data consist of two integrated-light spectra for each galaxy: one for the system's inner annulus and one for its outer annulus (as defined in Sect. \ref{sec:method}).  We fitted our observations to the SSP models using the Penalized Pixel-Fitting tool \citep[pPXF;][]{2004PASP..116..138C, 2017MNRAS.466..798C, 2023MNRAS.526.3273C}, with the  wavelength range limited to $4760~\mathrm{\AA} - 6800~\mathrm{\AA}$ to exclude sky lines in the redder part of the spectra. Additionally, we put masks centred on the telluric emission lines of [\ion{O}{i}]$\lambda 5577.4$, [\ion{O}{i}]6300.3, and [\ion{O}{i}]6363.7, \ion{Na}{i} $\lambda 5889.9$, and \ion{O}{i} $\lambda 6157.5$ \citep{2016A&A...591A.143G}.

The MUSE line spread function (LSF) is a non-linear function of wavelength. For our fitting range, the LSF varies from FWHM = 2.46 to 2.77~{\AA}\null.  In contrast, the sMILES SSP models have a constant FWHM = 2.51~{\AA}\null.  For regions of the sMILES spectra with a LSF FHWM $<$ 2.51~{\AA} we did not apply any convolution to the SSP models;  otherwise, we convolved the models with a kernel based on the LSF defined by \citet{2018A&A...611A..95W}:
\begin{equation}
    \begin{aligned}
        \mathrm{LSF}(\lambda) = 4.3459 - 3.6704 \times 10^{-4} \lambda - 6.0942 \times 10^{-9} \lambda^2 \\ + 2.8458 \times 10^{-12} \lambda^3
    \end{aligned}
\end{equation}
with the wavelength in \AA\null.
Then, to obtain the stellar population parameters, we applied the wild bootstrapping analysis \citep{davidson2008wild, 2023MNRAS.526.3273C}; a full description of this technique being applied to MUSE data using \texttt{pPXF} is given in \citet{2018MNRAS.480.1973K}. In brief, we first performed an initial fit to obtain the residuals at each pixel.  Then, we bootstrapped the data by resampling the residuals and refitting the spectrum 100 times. During this procedure, we applied a 12$^{th}$ order multiplicative polynomial to minimize any template mismatch and flux calibration errors, but we did not apply any additive polynomial correction, as that might bias the derived stellar population parameters \citep{2016A&A...591A.143G, 2018MNRAS.480.1973K}. Similarly, we did not apply any regularization for our measurements. Each of the 100 fits produced a luminosity-weighted measurements of the SSPs for different $[\mathrm{M/H}]$, $[\alpha/\mathrm{Fe}]$, and ages. We took the averaged weights of the 100 fits and their standard deviation to be our light-weighted results. To obtain the mass-weighted values, we adopted mass predictions for the SSP models from \citet{2015MNRAS.449.1177V} for the BaSTI isochrones \citep{2004ApJ...612..168P, 2006ApJ...642..797P}. We calculated the mass-weights for each of the 100 luminosity-weights, before obtaining the average mass-weighted results and their uncertainties.

\subsection{Star Formation Histories} \label{sec:SFH}

Based on the SSP mass-weights, we constructed the history of mass assembly of the stellar population. As our aim is to investigate the parent population contributing to the PNe at the bright-end of the PNLF, we needed to define a coarse age range where these objects are expected. Based on the simulations of \citet{2019ApJ...887...65V}, central stars with masses in the range of $0.53 - 0.59 \, \mathrm{M}_\odot$ are sufficient to reproduce the PNLF of the elliptical galaxy NGC 4697. This is further demonstrated by \citet{2025A&A...699A.371V}, who show which SSPs are capable of producing PNe at the bright cut-off (see their Fig.~10 for details). 

The lifetimes and luminosities of the progenitor stars of \oiii-bright PNe depend on their initial mass, metallicity and helium abundance;  thus it is important to constrain these variables. If we adopt the post-AGB models of \citet{2016A&A...588A..25M}, stars with main sequence lifetimes $>10$ Gyr will end up with central star masses of $<0.53 \, \mathrm{M}_\odot$ and be too faint to excite an $M^*$ planetary. PN hydrodynamical simulations suggested that to produce PNe at the PNLF cut-off, a central star mass $\gtrsim 0.58 \, \mathrm{M}_\odot$ is necessary \citep{2007A&A...473..467S, 2018NatAs...2..580G}. The observation of PNLF cut-off PNe in M31 showed that the central stars have masses of $0.57 -0.58 \, \mathrm{M}_\odot$ \citep{2022A&A...657A..71G, 2023A&A...674L...6M}; the masses correspond to main sequence lifetimes of $\sim 2-3$ Gyr. Recently, \citet{2025A&A...699A.371V} found that, depending on the IFMR, the $M^*$ PNe can be formed by stellar populations as old as 10 Gyr at solar metallicity. Based on these considerations, we defined three age bins: $t < 2$ Gyr, $2 \leq t \leq 10$ Gyr, and $t > 10$ Gyr to derive the stellar mass fractions and investigate the progenitor population of our PNe. Note that these definitions are model dependent on the populations models, in this case the sMILES SSPs.

\section{Simulation Analogues} 
\label{sec:simanalogue}

To interpret the relationship between the bright-end of the PNLF and underlying stellar populations, we apply the PICS model \citep[PNe In Cosmological Simulations;][]{2025A&A...699A.371V} in post-processing to the Magneticum Pathfinder\footnote{\url{www.magneticum.org}} simulation Box4 ultra-high resolution (uhr). This cosmological hydrodynamical simulation, which was run with \textsc{Gadget-3}, a modified version of the \textsc{Gadget-2} code \citep{2005MNRAS.364.1105S}, has a box side length of $68 \, \mathrm{Mpc}$, an average stellar particle mass of $\langle m_* \rangle = 1.8 \times 10^6 \, \mathrm{M}_\odot$ and a softening length of $\epsilon_* = 1 \, \mathrm{kpc}$; for more details about the implementation and setup, see \citet{2015ApJ...812...29T, 2025arXiv250401061D}. Galaxies and their substructures are identified with the subhalo finder \textsc{SubFind} \citep{2001MNRAS.328..726S, 2009MNRAS.399..497D}. In the simulation volume roughly 700 galaxies with stellar masses above $M_\bigstar \approx 2 \times 10^{10} \, \mathrm{M}_\odot$ are formed by $z = 0$. The galaxies of Box4 (uhr) have been analysed in numerous studies and have been shown to agree well with observations in a multitude of properties. An overview of these studies can be found in Sect.~2.1 of \citet{2024A&A...690A.206V}.  One caveat to note about the Box4 simulation is that the run only reaches $z \approx 0.066$, which corresponds to a look-back time of around $0.88\,\mathrm{Gyr}$. For this reason, the stellar ages reach a maximum of $12.9\,\mathrm{Gyr}$.

We select six simulated analogue galaxies for each of our five observed objects. Due to the limited size of the simulation box, we are limited to a low number of galaxies. We find that six galaxies give a compromise between not bound to a single object and having a sample within our selection parameters. For the selection, we do this matching, as best as possible: morphological type (here, either elliptical and lenticular), effective radius (as reported by \citet{2017MNRAS.464.4611F} and \citet{2019A&A...627A.136I} assuming the distances of \citet{2024ApJS..271...40J}), and $g - r$ colour \citep{1979ApJS...39...61P, 2019A&A...623A...1I}\footnote{When necessary, colour conversion into $g - r$ is done using  https://classic.sdss.org/dr5/algorithms/sdssUBVRITransform.php}. Specifically, we began by discarding all simulated galaxies classified as late-type by their dynamical morphology indicator \citep[the $b$ value;][]{2015ApJ...812...29T} according to the classification scheme employed by \citet{2024A&A...686A.182V}. Next, we excluded systems whose effective radii differed by more than 30\% from their observed counterparts, or more than $\pm 0.4$ from the observed $g-r$ colour.  Finally, we identified the best six galaxies in the Magneticum simulation via the metric
\begin{equation}
    m = \left( \frac{R_{e,\mathrm{sim}} - R_{e,\mathrm{obs}}}{0.3 R_{e,\mathrm{obs}}} \right)^2 + \left( \frac{(g-r)_\mathrm{sim} - (g-r)_\mathrm{obs}}{0.4} \right)^2,
\end{equation}
where the galaxies with the lowest values of $m$ were taken as analogues. Only duplicate analogues for two different observed galaxies are skipped.

For each of the galaxy analogues, we ran the fiducial PICS model on each stellar particle belonging to its respective galaxy; this results in set of PNe, with a position and a $M_{5007}$ magnitudes.  The PICS procedure is as follows. As input, PICS takes a single stellar population with a given age, metallicity mass fraction, total mass and initial mass function (IMF\null). From the age, the number of stars now entering the post-AGB phase are determined through the lifetime function \citep{2016A&A...588A..25M},  the total mass and IMF\null. Each post-AGB object is then assigned a core mass through the \citet{2016A&A...588A..25M} initial-to-final mass relation (IFMR), and then a stellar luminosity and effective temperature using the \citet{2016A&A...588A..25M} evolutionary tracks and a randomly drawn post-AGB age. Note that the lifetime function, IFMR, and post-AGB tracks are all metallicity-dependent. Based on the central star properties and post-AGB age, the intrinsic $M_{5007}$ magnitude is obtained from the nebular model of \citet{2019ApJ...887...65V} corrected for metallicity according to \citet{1992ApJ...389...27D}. Finally, the actually measured magnitude is obtained by applying the circumnebular extinction relation by \citet{2025ApJ...983..129J}.  For more details on the PICS model, see \citet{2025A&A...699A.371V}.

To obtain the PNe in analogueous radial bins to the observations, we used the values given in Table~\ref{tab:my_label} for $a_1$, $a_2$, and $a_3$ and converted them to kpc according to the respective PNLF distances.  We then viewed the simulated galaxy from a random direction, and, using the ellipticities listed in Table~\ref{tab:my_label}, defined its inner and outer annulus in a manner similar to that of the observed galaxy.  The PNe found in the simulation's projected regions then serve as the comparison sample for the observed PNe.  

The analogue galaxies function as hosts of PN populations with self-consistent cosmological star formation histories, which we can compare to observations.  This allows us to gain a deeper understanding of the underlying processes and how galaxy properties are related to their PN population.  This, in turn can be compared to the observational findings.  However, we note that none of the simulated galaxies are exact replicas of the observed systems nor were they produced under constraints to reproduce the program galaxies.  Thus, the comparisons may not be perfect.


\section{Results} \label{sec:results}

\subsection{PNLF and Stellar Populations}

\subsubsection{NGC 1052}

\begin{figure}[]
    \centering
    \includegraphics[width=0.9\hsize]{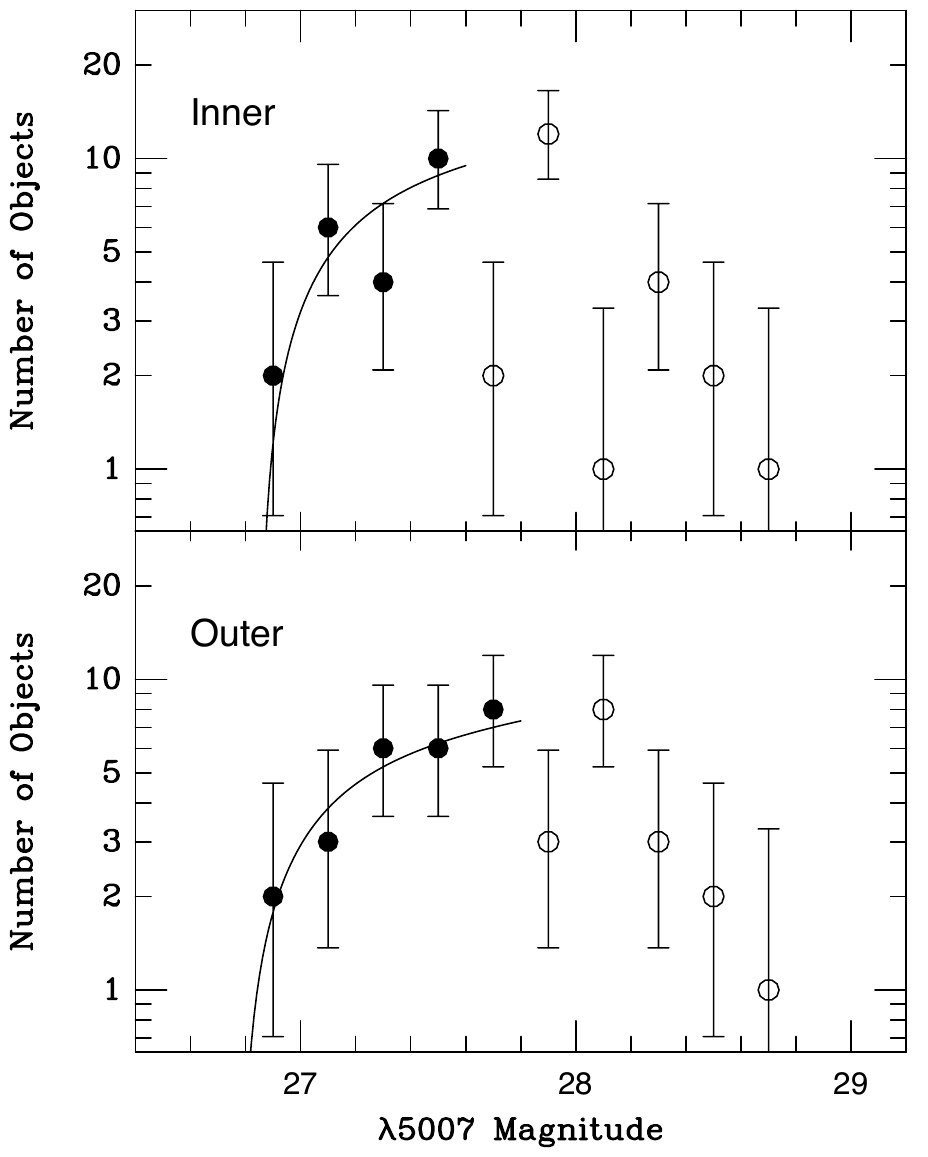}
    \caption{PNLFs of the inner and outer region of NGC 1052, binned into 0.2~mag intervals.  Solid and open points represent data brighter than, and fainter than the completeness limit, respectively.  The curves show the analytic PNLF, shifted using the most likely apparent distance modulus and normalization.}
    \label{fig:PNLF_NGC1052}
\end{figure}

NGC 1052 is a bright elliptical galaxy with a LINER-type nucleus which is known to have X-ray variability \citep{2015ApJ...801...42D, 2020MNRAS.491...29O}. A total of 86 PNe were identified in Paper II, giving us 43 PNe per ellipse region. To measure the distance modulus, we adopted the $R$-band surface brightness photometry from \citet{1987AJ.....94.1508J}, used the velocity dispersion data of \citet{1990ApJ...361...78B}, and assumed a population $(V-R) = 0.93$ colour \citep{1979ApJS...39...61P} with $E(B-V) = 0.023$ \citep{2011ApJ...737..103S}.  For the inner annulus, we estimated the limiting PN magnitude for 90\% completeness to be $m_{5007}$ = 27.6, while in the outer annulus, the galaxy's lower surface brightness enabled measurements to $m_{5007} = 27.8$. The PNLFs are shown in Fig. \ref{fig:PNLF_NGC1052}. 

\begin{table}[]
    \centering
    \caption{Fit parameters for NGC 1052.}
    \begin{tabular}{c c c}
    \hline
        PNLF & $0.45 <R_e<1.30$ & $1.30  < R_e < 2.74$\\
    \hline
        $m^*$ & $26.77^{+0.05}_{-0.10}$ & $26.70^{+0.05}_{-0.11}$ \\
        $\alpha_{0.5} \: [\times 10^{-9} \: \mathrm{PN/L}_\odot]$ & $1.2^{+0.5}_{-0.3}$ & $1.4^{+0.5}_{-0.4}$ \\

    \hline
        Stellar populations &  & \\
    \hline
        Age$_{\mathrm{light}}$ [Gyr] & 12.46 $\pm$ 1.16 & 12.02 $\pm$ 1.60 \\
        Age$_{\mathrm{mass}}$ [Gyr]& 12.89 $\pm$ 0.99 & 12.48 $\pm$ 1.27 \\
        $[\mathrm{M/H}]_{\mathrm{light}}$ & 0.11 $\pm$ 0.03 & 0.00 $\pm$ 0.04 \\
        $[\mathrm{M/H}]_{\mathrm{mass}}$ & 0.13 $\pm$ 0.03 & 0.02 $\pm$ 0.04 \\
        $[\alpha/\mathrm{Fe}]_{\mathrm{light}}$ & 0.22 $\pm$ 0.02 & 0.27 $\pm$ 0.03 \\
        $[\alpha/\mathrm{Fe}]_{\mathrm{mass}}$ & 0.20 $\pm$ 0.06 & 0.25 $\pm$ 0.06 \\
    \hline
    Age fractions &  &  \\
    \hline
        $t < 2$ Gyr [\%]  & 1.05 $\pm$ 1.44 & 1.04 $\pm$ 1.46 \\
        $2 \leq t \leq 10$ Gyr [\%]  & 4.72 $\pm$ 9.59 & 7.91 $\pm$ 12.24 \\
        $t > 10$ Gyr [\%]  & 94.23 $\pm$ 10.05 & 91.05 $\pm$ 12.15 \\
    \hline
    \end{tabular}
    \label{tab:ngc1052_fit}
\end{table}

The fitted parameters are shown in Table \ref{tab:ngc1052_fit}. The difference between the PNLF cut-off between the two regions is $\Delta m^* = 0.07^{+0.07}_{-0.15}$. The PN density in both annuli is similar with $\Delta \alpha_{0.5} = 0.2^{+0.7}_{-0.5} \times 10^{-9} \: \mathrm{PN/L}_\odot$. Both regions are very old, with average weighted ages greater than 12 Gyr. The inner region appears more metal-rich than the outer by $\sim$ 0.1 dex, while the difference in $[\alpha/\mathrm{Fe}]$ is not as significant. 
The errors in the age fraction  are generally high due to the uncertainties in the fitting process. Regardless, our results are generally in agreement with previous studies of NGC\,1052 that found an older and more metallic population in the inner part of the galaxy \citep{2001MNRAS.324.1087R, 2007A&A...469...89M, 2019MNRAS.482.5211D}.

\subsubsection{NGC 1351}

NGC 1351 is a peculiar lenticular galaxy \citep[Hubble type SA0;][]
{1991rc3..book.....D} $1\fdg 7$ from the centre of the Fornax Cluster.  In Paper II, 102 PNe were found, but as explained in Sect. \ref{sec:data}, for this differential analysis we only use the PN detections in the MUSE pointing centred on the nucleus.  This gives us a total sample of 86 PNe. Our  surface photometry data came from the $B$-band observations of \citet{1991ApJS...76.1067D}, which was then converted to $V$ using $(B-V) = 0.90$ \citep{1989ApJS...69..763F} and $E(B-V) = 0.016$ \citep{2011ApJ...737..103S}.  These data yield $V = 14.23$ for the light contained in our inner annulus and $V = 14.65$ for light of the outer annulus.  The line-of-sight velocity dispersion underlying each PN was estimated from the data presented by \citet{1995A&A...296..319D}. 

\begin{figure}[]
    \centering
    \includegraphics[width=0.9\hsize]{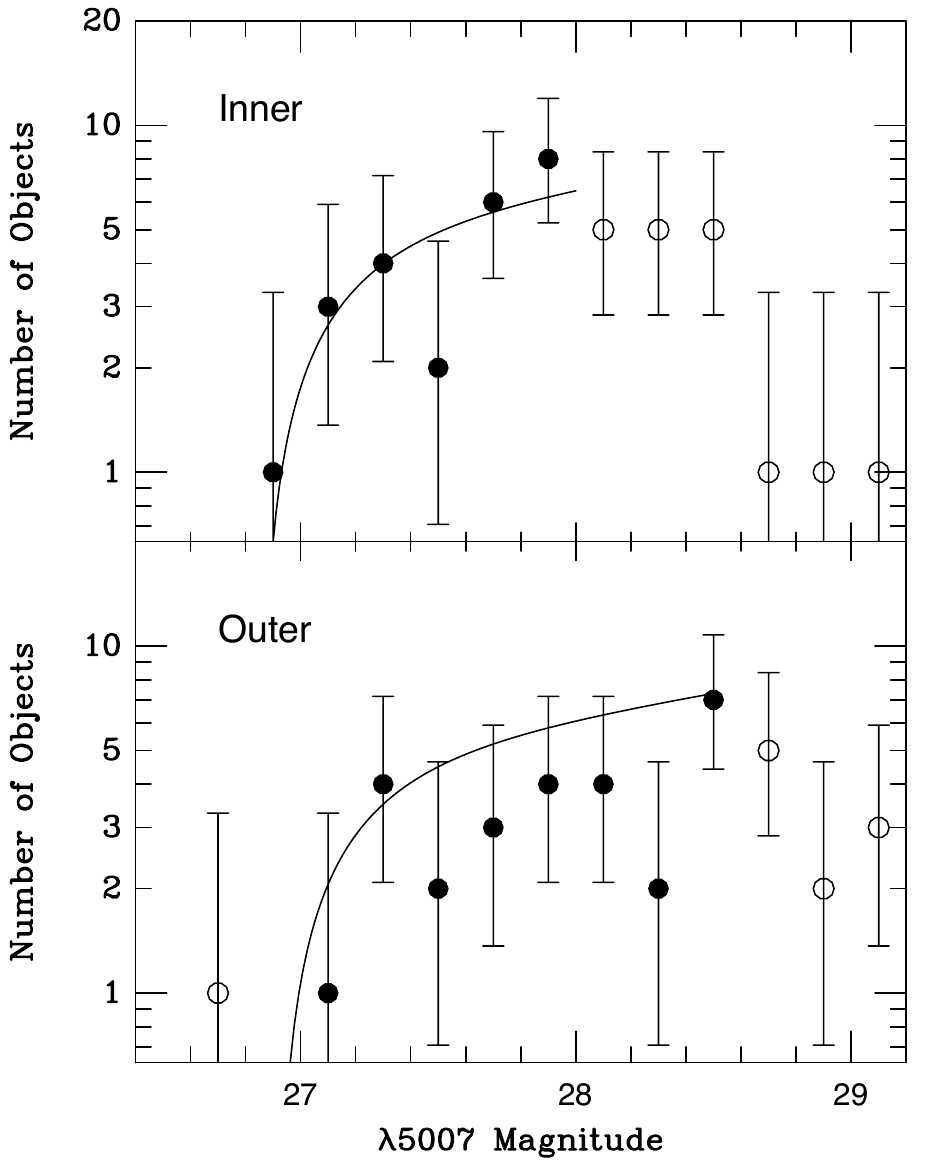}
    \caption{PNLFs of the inner and outer regions of NGC 1351, binned into 0.2~mag intervals. Solid and open points represent data brighter than and fainter than, the completeness limit, respectively. One over-luminous PN was excluded from the outer PN sample; that object is also shown as an open circle.  The curves show the analytic PNLF, shifted using the most likely apparent distance modulus and normalization.}
    \label{fig:PNLF_NGC1351}
\end{figure}

We assume our measurements to be complete to $m_{5007} = 28.0$ in the galaxy's high surface-brightness inner annulus, and $m_{5007} = 28.5$ at the inner and outer annulus; our exact choice of limiting magnitude has no effect on the derived distance and only a minor effect on the PN density. The PNLF plots are shown in Fig. \ref{fig:PNLF_NGC1351} and the parameters from our analysis are presented in Table \ref{tab:ngc1351_fit}. Note that we exclude one, apparently over-luminous PN in the outer annulus of the galaxy; this object is fully discussed in Paper~II. 

\begin{table}[]
    \centering
    \caption{Fit parameters for NGC 1351.}
    \begin{tabular}{c c c}
    \hline 
        PNLF & $0.19 <R_e<0.90$ & $0.90  < R_e < 2.18$\\
    \hline 
        $m^*$ & $26.81^{+0.06}_{-0.13}$ & $26.87^{+0.07}_{-0.17}$ \\
        $\alpha_{0.5} \: [\times 10^{-9} \: \mathrm{PN/L}_\odot]$ & $4.9^{+1.9}_{-1.3}$ & $4.3^{+1.6}_{-1.1}$ \\
    
    \hline 
        Stellar populations &  & \\
    \hline
    
        Age$_{\mathrm{light}}$ [Gyr] & 11.08 $\pm$ 0.71 & 9.29 $\pm$ 1.42 \\
        Age$_{\mathrm{mass}}$ [Gyr]& 11.25 $\pm$ 0.71 & 9.17 $\pm$ 1.35 \\
        $[\mathrm{M/H}]_{\mathrm{light}}$ & -0.29 $\pm$ 0.03 & -0.41 $\pm$ 0.05 \\
        $[\mathrm{M/H}]_{\mathrm{mass}}$ & -0.30 $\pm$ 0.07 & -0.43 $\pm$ 0.12 \\
        $[\alpha/\mathrm{Fe}]_{\mathrm{light}}$ & 0.31 $\pm$ 0.02 & 0.28 $\pm$ 0.05 \\
        $[\alpha/\mathrm{Fe}]_{\mathrm{mass}}$ & 0.27 $\pm$ 0.05 & 0.25 $\pm$ 0.07 \\
    \hline
    
    Age fractions &  &  \\
    \hline
    
        $t < 2$ Gyr [\%]  & 0.26 $\pm$ 0.47 & 2.38 $\pm$ 3.00 \\
        $2 \leq t \leq 10$ Gyr [\%]  & 31.47 $\pm$ 15.18 & 56.18 $\pm$ 24.47 \\
        $t > 10$ Gyr [\%]  & 68.28 $\pm$ 15.05 & 41.44 $\pm$ 23.83 \\
    \hline
    
    \end{tabular}
    \label{tab:ngc1351_fit}
\end{table}

Between the two regions, the difference in the PNLF cut-off is an insignificant $\Delta m^* = 0.06 ^{+0.09}_{-0.21}$. Similarly, the PN densities within the two regions are also consistent, with $\Delta \alpha_{0.5} = 0.6^{+2.5}_{-1.6} \times 10^{-9} \: \mathrm{PN/L}_\odot$. On average, both regions are old, with mean ages $\sim 11$ Gyr and $\sim 9$ Gyr for the inner and outer region, respectively. The total metallicity in the inner annulus is $[\mathrm{M/H}] \sim -0.3$, which is $\sim 0.1$ dex more metal rich than the outer annulus but within the uncertainties. The $[\alpha/\mathrm{Fe}]$ of the two regions are not significantly different. Compared with the stellar population calculation of \citet{2019A&A...627A.136I}, we have younger ages by $\sim 1-2$ Gyr and lower total metallicity. We argue that this is due to the model dependence of the SSP fitting, as the latter study employed the older MILES SSPs; support for this in given in Appendix \ref{appendix:SSP_comparison}.  Nevertheless, the overall radial trends for our stellar population parameters are in an agreement.

\subsubsection{NGC 1380}

NGC 1380 is another lenticular (SA0) galaxy in the Fornax cluster. It was one of the benchmark galaxy for the demonstration of the DELF method of MUSE reduction (Paper~I), and has been the subject of numerous analyses, in part because of its well-observed Type~Ia supernova 1992A\null.  In particular, studies of the galaxy's stellar initial mass function, kinematics, and gas components have been performed as part of the Fornax3D project \citep{2018A&A...616A.121S, 2019A&A...627A.136I, 2019A&A...626A.124M, 2019A&A...622A..89V}. 

In Paper I, 166 PNe were identified in three MUSE pointings; 112 of these are present in the nucleus data cube.  Our PNLF analysis uses the $ugri$ surface photometry of \citet{2019A&A...623A...1I}, which was converted to the $V$-band using the transformations given by \citet{lupton2005}, and a foreground extinction value of $E(B-V) = 0.015$ \citep{2011ApJ...737..103S}.  This yielded $V = 11.88$ for the galaxy's inner annulus and $V = 11.92$ for our outer annulus.  The line-of-sight kinematics underlying each PN were found from \citet{2019A&A...627A.136I}. 

\begin{figure}[]
    \centering
    \includegraphics[width=0.9\hsize]{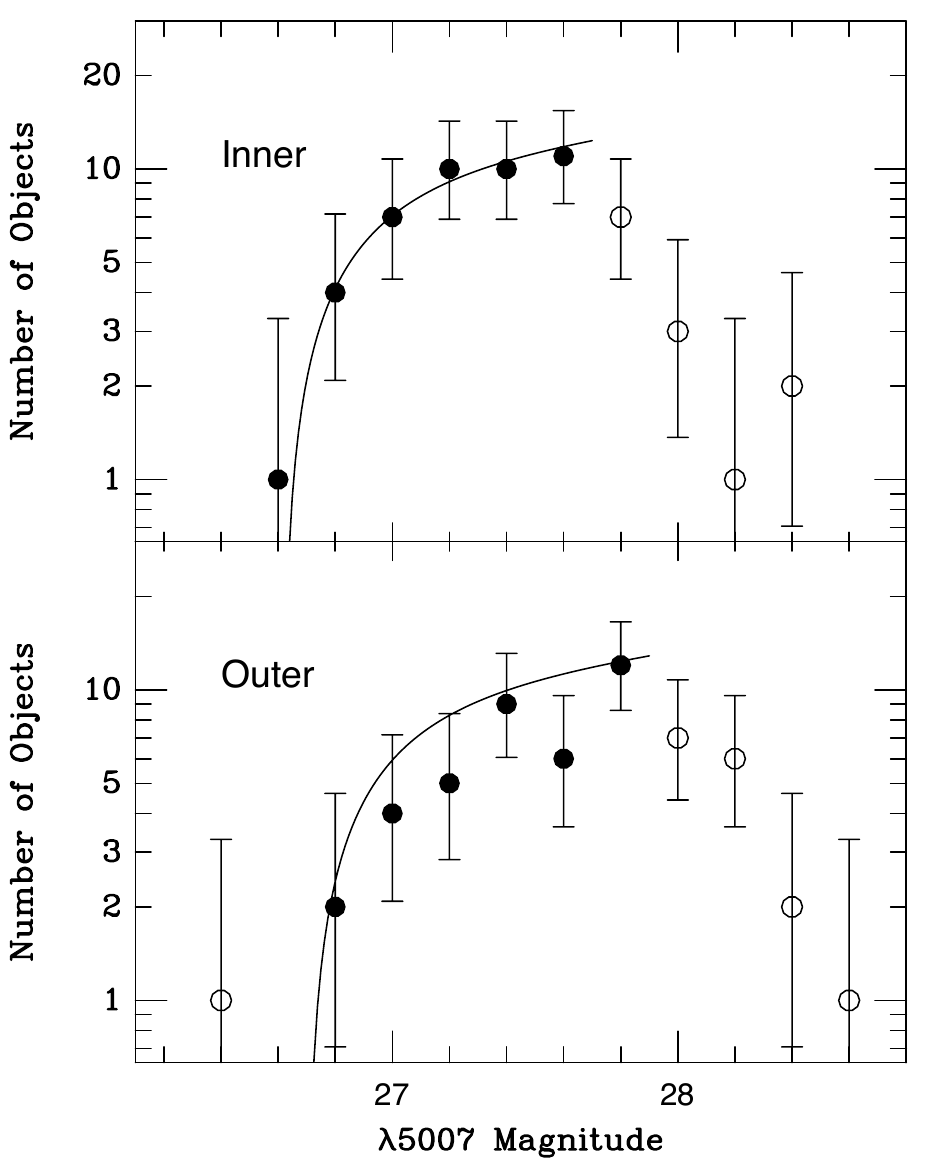}
    \caption{PNLFs of the inner and outer region of NGC 1380, binned into 0.2~mag intervals. Solid and open points represent PN bins brighter than, and fainter than the completeness limit, respectively. One over-luminous PN was excluded from the outer PNLF sample, and is shown with an open circle.  The curves show the analytic PNLF, shifted using the most likely apparent distance modulus and normalization.}
    \label{fig:PNLF_NGC1380}
\end{figure}

\begin{table}[]
    \centering
    \caption{Fit parameters for NGC 1380.}
    \begin{tabular}{c c c}
    \hline 
        PNLF & $0.12 <R_e<0.36$ & $0.36  < R_e < 0.73$\\
    \hline 
        $m^*$ & $26.56^{+0.06}_{-0.09}$ & $26.65^{+0.07}_{-0.10}$ \\
        $\alpha_{0.5} \: [\times 10^{-9} \: \mathrm{PN/L}_\odot]$ & $1.4^{+0.4}_{-0.3}$ & $1.0^{+0.3}_{-0.2}$ \\
    
    \hline 
        Stellar populations &  & \\
    \hline
    
        Age$_{\mathrm{light}}$ [Gyr] & 12.49 $\pm$ 0.67 & 12.13 $\pm$ 0.69 \\
        Age$_{\mathrm{mass}}$ [Gyr]& 12.85 $\pm$ 0.53 & 12.42 $\pm$ 0.63 \\
        $[\mathrm{M/H}]_{\mathrm{light}}$ & -0.06 $\pm$ 0.02 & -0.11 $\pm$ 0.03 \\
        $[\mathrm{M/H}]_{\mathrm{mass}}$ & 0.00 $\pm$ 0.05 & -0.07 $\pm$ 0.04 \\
        $[\alpha/\mathrm{Fe}]_{\mathrm{light}}$ & 0.25 $\pm$ 0.02 & 0.24 $\pm$ 0.02 \\
        $[\alpha/\mathrm{Fe}]_{\mathrm{mass}}$ & 0.21 $\pm$ 0.03 & 0.20 $\pm$ 0.04 \\
    \hline
    
    Age fractions &  &  \\
    \hline
    
        $t < 2$ Gyr [\%]  & 0.84 $\pm$ 0.75 & 0.83 $\pm$ 0.87 \\
        $2 \leq t \leq 10$ Gyr [\%]  & 5.33 $\pm$ 5.77 & 9.06 $\pm$ 7.54 \\
        $t > 10$ Gyr [\%]  & 93.83 $\pm$ 5.80 & 90.11 $\pm$ 7.51 \\
    \hline
    
    \end{tabular}
    \label{tab:ngc1380_fit}
\end{table} 

The luminosity functions are shown in Fig. \ref{fig:PNLF_NGC1380}. We assume a limiting magnitude of $m_{5007} = 27.7$ and $m_{5007} = 27.9$ for the inner and outer annulus, respectively. The PNLF fit and stellar population parameters are presented in Table \ref{tab:ngc1380_fit}. Once again, we excluded one PN candidate that is over-luminous compared to the rest of the population in the outer annulus; see Paper~I for details. 

The PNe of the inner annulus of NGC\,1380 exhibits a brighter PNLF cut-off ($\Delta m^* = 0.09 ^{+0.09}_{-0.13}$) and a lower PN density ($\Delta \alpha_{0.5} = 0.4^{+0.5}_{-0.4} \times 10^{-9} \: \mathrm{PN/L}_\odot$) than their outer annulus counterparts, but neither offset is significant.   The average stellar age is $\sim 12-13$ Gyr. The inner annulus exhibits solar metallicity, while the stars in the outer annulus is subsolar with $[\mathrm{M/H}] \sim -0.10$.  There is no significant difference in $[\alpha/\mathrm{Fe}]$ between the regions. Our values are in agreement with the measurements by \citet{2019A&A...627A.136I} and \citet{2019A&A...626A.124M}. With respect to the latter, we note that they have explored the variability of the initial mass function in this galaxy, which we did not take account for.

\subsubsection{NGC 1399}

NGC 1399 is the dominant cD/elliptical galaxy of the Fornax cluster, and, due to its mass and MUSE coverage, it has more detected PNe than any other galaxy in Paper~II\null. The first observations of PN in the elliptical galaxy NGC 1399 date back to the multi-object spectroscopy with the ESO NTT by \citet{1994Msngr..76...40A}. Follow-up observations with the ESO FORS spectrograph in counter-dispersed slit-less spectroscopy detected 146 PNe covering a radial range in the outer halo from $100\arcsec$ to $500\arcsec$ \citep{2010A&A...518A..44M}. 

\begin{figure}[]
    \centering
    \includegraphics[width=0.9\hsize]{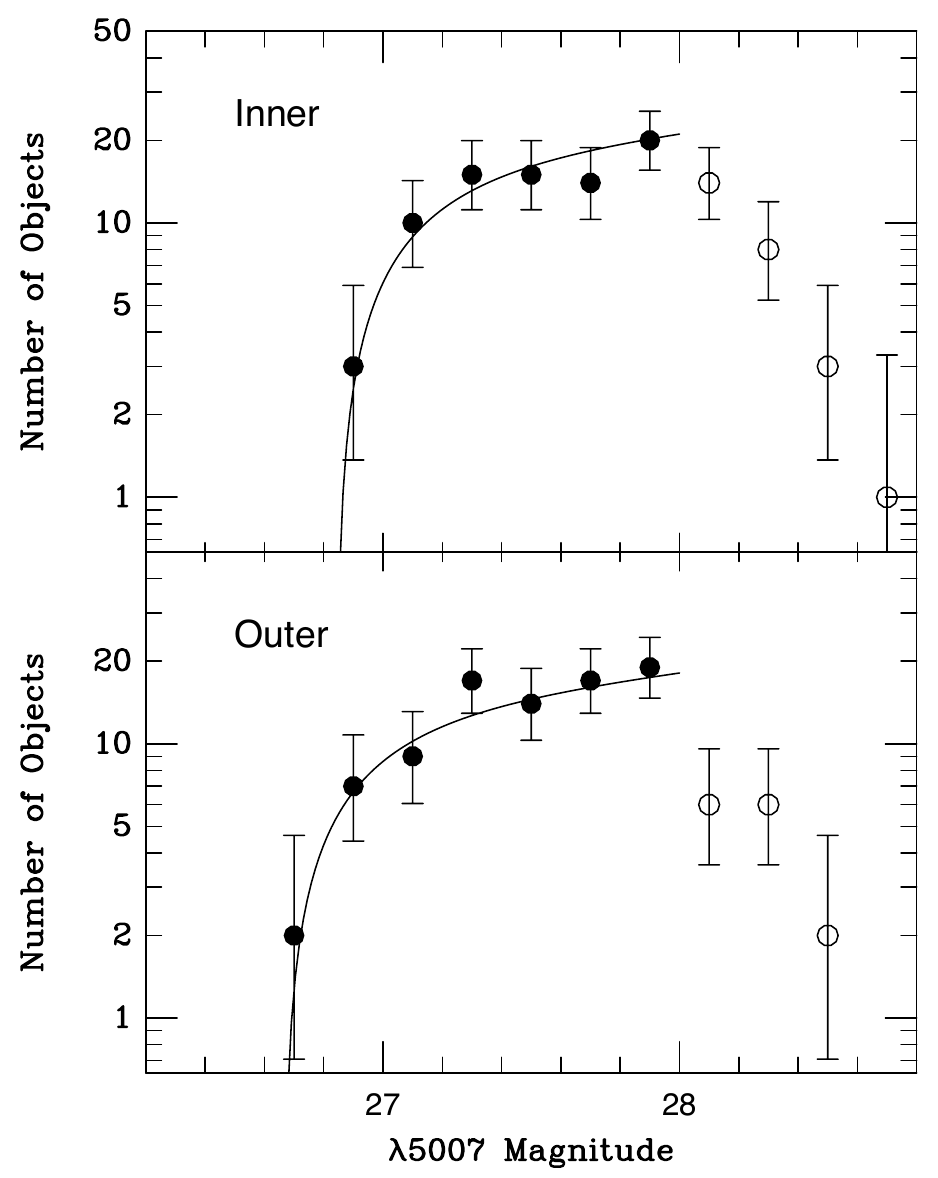}
    \caption{PNLFs of the inner and outer region of NGC 1399, binned into 0.2~mag intervals. Solid and open data points represent data brighter than, and fainter than, the completeness limit, respectively.  The curves show the analytic PNLF, shifted using the most likely apparent distance modulus and normalization.}
    \label{fig:PNLF_NGC1399}
\end{figure}

For the current investigation of the PNLF properties we restrict ourselves to the central regions of NGC 1399 sampled with MUSE. As in Paper II, we excluded all PNe within $15\arcsec$ of the galaxy's nucleus, as detections in this region are compromised by the bright, rapidly varying background. The galaxy's surface brightness distribution has been measured several times at different depths and scales \citep{1989AJ.....98..538F, 1994A&AS..106..199C, 2016ApJ...820...42I}; in our analysis, we combined these data with an assumed $E(B-V) = 0.011$ \citep{2011ApJ...737..103S} to create the galaxy's $V$-band surface brightness profile.  The line-of-sight velocity dispersion is adopted from \citep{2000AJ....119..153S}. Outside of this region, we used a common limiting magnitude of $m_{5007} = 28.0$.  Using the galaxy's $V$-band surface brightness distribution, we calculate that our inner annulus of PNe contains $V = 11.13$ of total light, while the outer annulus encompasses $V=11.40$. Our measured luminosity functions are shown in Fig. \ref{fig:PNLF_NGC1399}. 

\begin{table}[]
    \centering
    \caption{Fit parameters for NGC 1399.}
    \begin{tabular}{c c c}
    \hline 
        PNLF & $0.04 <R_e<0.13$ & $0.13  < R_e < 0.32$\\
    \hline 
        $m^*$ & $26.80^{+0.05}_{-0.07}$ & $26.62^{+0.06}_{-0.07}$ \\
        $\alpha_{0.5} \: [\times 10^{-9} \: \mathrm{PN/L}_\odot]$ & $0.9^{+0.2}_{-0.1}$ & $1.2^{+0.2}_{-0.2}$ \\
    
    \hline 
        Stellar populations &  & \\
    \hline
    
        Age$_{\mathrm{light}}$ [Gyr] & 11.59 $\pm$ 0.96 & 12.18 $\pm$ 0.90 \\
        Age$_{\mathrm{mass}}$ [Gyr]& 12.25 $\pm$ 0.81 & 12.65 $\pm$ 0.74 \\
        $[\mathrm{M/H}]_{\mathrm{light}}$ & 0.08 $\pm$ 0.03 & 0.09 $\pm$ 0.04 \\
        $[\mathrm{M/H}]_{\mathrm{mass}}$ & 0.10 $\pm$ 0.03 & 0.11 $\pm$ 0.03 \\
        $[\alpha/\mathrm{Fe}]_{\mathrm{light}}$ & 0.23 $\pm$ 0.02 & 0.27 $\pm$ 0.03 \\
        $[\alpha/\mathrm{Fe}]_{\mathrm{mass}}$ & 0.20 $\pm$ 0.04 & 0.26 $\pm$ 0.05 \\
    \hline
    
    Age fractions &  &  \\
    \hline
    
        $t < 2$ Gyr [\%]  & 2.79 $\pm$ 1.30 & 1.63 $\pm$ 1.45 \\
        $2 \leq t \leq 10$ Gyr [\%]  & 1.73 $\pm$ 4.34 & 1.75 $\pm$ 4.05 \\
        $t > 10$ Gyr [\%]  & 95.48 $\pm$ 4.49 & 96.62 $\pm$ 4.06 \\
    \hline
    
    \end{tabular}
    \label{tab:ngc1399_fit}
\end{table}

The fitted parameters for the PNLF and stellar populations are given in Table \ref{tab:ngc1399_fit}.  Compared to the PNe in the outer annulus of the galaxy, the PNe in the inner annulus have a PN density that is lower by $\Delta \alpha_{0.5} = 0.3^{+0.3}_{-0.2} \times 10^{-9} \: \mathrm{PN/L}_\odot$, and a cut-off magnitude that is fainter by $\Delta m^* = 0.18 ^{+0.08}_{-0.10}$; both offsets are intriguing, but not definitive.  In contrast, our stellar population fits show that both annuli have similar ages and $\mathrm{[M/H]}$ values.  The $[\alpha/\mathrm{Fe}]$ estimate for the outer annulus is larger than that of the inner region, but still within their uncertainties. These stellar population trends are consistent with the findings of \citet{2018MNRAS.479.2443V}. 

\subsubsection{NGC 1404}

This galaxy is the second brightest galaxy in the Fornax cluster, located in vicinity to NGC 1399. 
The counter-dispersed slit-less spectroscopy observations of NGC 1404 by \citet{2010A&A...518A..44M} provided 23 PNe associated with the light of NGC 1404. On the basis of a Gaussian Mixture Model decomposition of the line-of-sight (LOS) velocity distribution, it shows the different LOS systemic velocity between NGC~1404 ($1933 $kms$^{-1}$) and NGC~1399 ($1425 $kms$^{-1}$).

Again, for the current investigation of the PNLF properties, we restrict ourselves to the central regions of NGC 1404 sampled with MUSE. In Paper II, a total of 126 PN candidates were identified, with 107 PNe in the MUSE pointing that we study here. This gives us 54 and 53 PN samples for the inner and the outer annulus, respectively.  
 
\begin{figure}[]
    \centering
    \includegraphics[width=0.9\hsize]{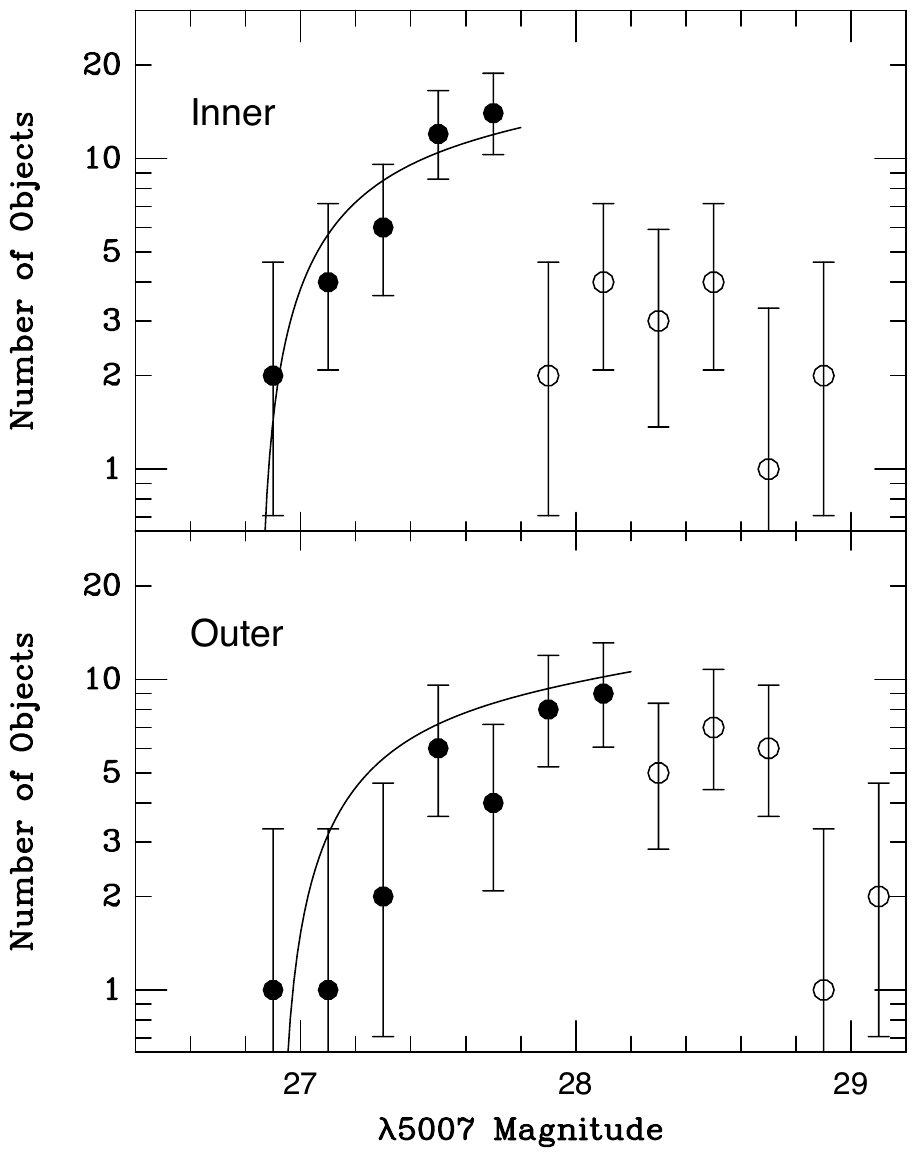}
    \caption{PNLFs of the inner and outer region of NGC 1404, binned into 0.2~mag intervals. Solid and open data points represent data brighter than, and fainter than, the completeness limit, respectively.  The curves show the analytic PNLF, shifted using the most likely apparent distance modulus and normalization.}
    \label{fig:PNLF_NGC1404}
\end{figure}

To derive the PNLF, we employed the $R$-band surface photometry of \citet{1989AJ.....98..538F}, converted it to the $V$-band using a colour of $(V-R) = 0.38$ \citep{2009ApJ...703.1569M}, and adopted a foreground extinction of 
$E(B-V)= 0.010$ \citep{2011ApJ...737..103S}.  Our information about the line-of-sight velocity dispersion came from \citet{1995A&A...296..319D} and \citet{2019A&A...623A...1I}. For the inner annulus, we calculated a $V$-band magnitude of $V = 10.81$ and assumed a completeness limit of $m_{5007} = 27.8$. For the lower surface brightness outer annulus, $V = 12.17$ and a completeness limit reaches $m_{5007} = 28.2$. The PNLF of the two sub-populations in NGC 1404 can be seen in Fig. \ref{fig:PNLF_NGC1404}. The fitted parameter for the PNLF cut-off and the stellar population are presented in Table \ref{tab:ngc1404_fit}. 

\begin{table}[]
    \centering
    \caption{Fit parameters for NGC 1404.}
    \begin{tabular}{c c c}
    \hline 
        PNLF & $0.03 <R_e<0.13$ & $0.13  < R_e < 0.29$\\
    \hline 
        $m^*$ & $26.82^{+0.05}_{-0.09}$ & $26.89^{+0.07}_{-0.12}$ \\
        $\alpha_{0.5} \: [\times 10^{-9} \: \mathrm{PN/L}_\odot]$ & $0.4^{+0.1}_{-0.1}$ & $0.7^{+0.3}_{-0.2}$ \\
    
    \hline 
        Stellar populations &  & \\
    \hline
    
        Age$_{\mathrm{light}}$ [Gyr] & 12.18 $\pm$ 0.86 & 12.99 $\pm$ 0.70 \\
        Age$_{\mathrm{mass}}$ [Gyr]& 12.69 $\pm$ 0.67 & 13.29 $\pm$ 0.48 \\
        $[\mathrm{M/H}]_{\mathrm{light}}$ & 0.13 $\pm$ 0.02 & -0.04 $\pm$ 0.03 \\
        $[\mathrm{M/H}]_{\mathrm{mass}}$ & 0.15 $\pm$ 0.02 & 0.00 $\pm$ 0.03 \\
        $[\alpha/\mathrm{Fe}]_{\mathrm{light}}$ & 0.16 $\pm$ 0.01 & 0.25 $\pm$ 0.02 \\
        $[\alpha/\mathrm{Fe}]_{\mathrm{mass}}$ & 0.15 $\pm$ 0.02 & 0.23 $\pm$ 0.04 \\
    \hline
    
    Age fractions &  &  \\
    \hline
    
        $t < 2$ Gyr [\%]  & 2.07 $\pm$ 0.99 & 0.39 $\pm$ 0.49 \\
        $2 \leq t \leq 10$ Gyr [\%]  & 0.29 $\pm$ 1.09 & 2.45 $\pm$ 3.59 \\
        $t > 10$ Gyr [\%]  & 97.64 $\pm$ 1.40 & 97.16 $\pm$ 3.64 \\
    \hline
    
    \end{tabular}
    \label{tab:ngc1404_fit}
\end{table} 

We see no significant shift in the PNLF cut-off magnitude between the two populations with $\Delta m^* = 0.07 ^{+0.09}_{-0.15}$. The PN density in the outer annulus is larger by $\Delta \alpha_{0.5} = 0.3^{+0.3}_{-0.2} \times 10^{-9} \: \mathrm{PN/L}_\odot$, but, as indicated by the uncertainties, this difference is, at best, marginally significant.  The stellar population trends for the age and abundances of each regions are also in agreement with the results of \citet{2019A&A...627A.136I}. Both annuli exhibit an old population with the average age of $12 - 13$ Gyr. This is also shown by the large fraction ($\sim 97\%$) of old stars in the galaxy's star formation history. The remaining fraction of $\sim 2\%$ in the inner annulus is dominated by stars younger than 2 Gyr, while in the outer annulus, this population has intermediate ages between 2-10 Gyr.

\subsubsection{Observation Summary}

We have measured the PNLFs and stellar populations in ten regions within five galaxies. Although the $\Delta m^*$ and $\Delta\alpha_{0.5}$ between regions within a single galaxy may not be significant, a combinations of all the result will allow us to investigate whether there are significant correlations between the PNLF parameters and the stellar population parameters. We discuss this in details in Sect. \ref{sec:Mstar} and Sect. \ref{sec:apar}. 

\subsection{Simulated Analogue Galaxies}

\begin{figure*}[]
    \centering
    \includegraphics[width=0.85\linewidth]{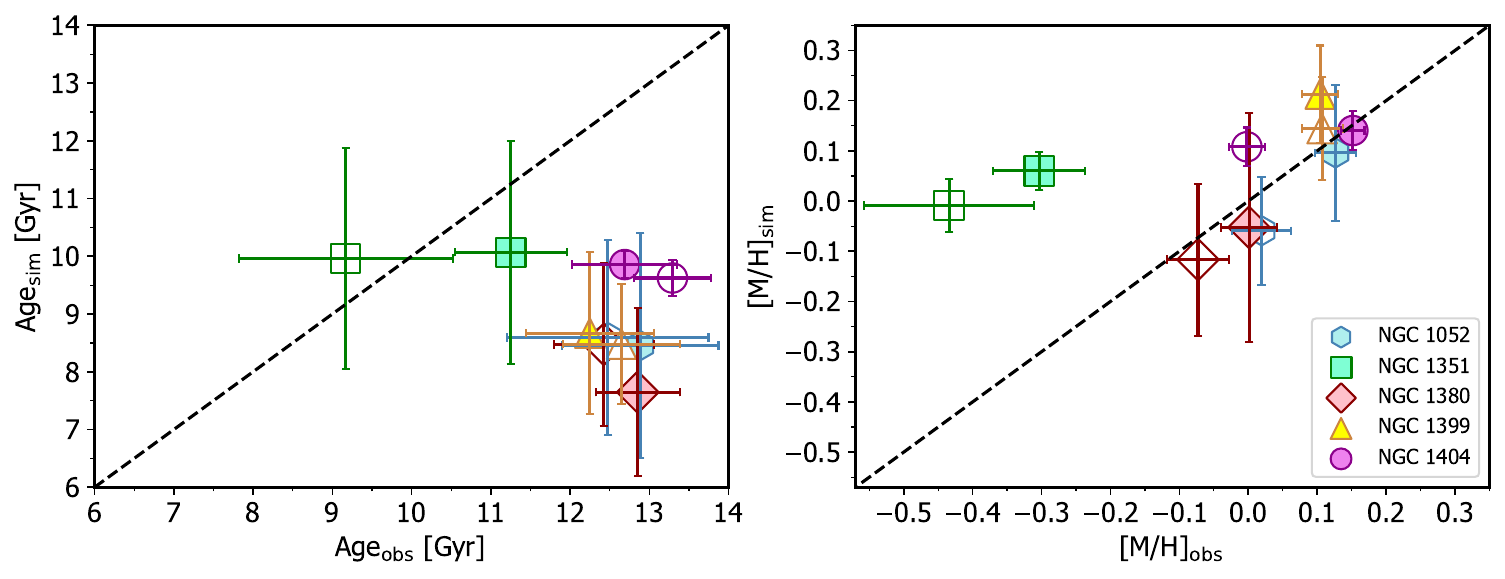}
    \caption{Comparison of the stellar population age (left) and total metallicity (right) between the observations and the simulations. Solid and open markers indicate the inner and outer annulus, respectively. The dashed line indicates one-to-one comparison.}
    \label{fig:observation_simulation_ssp}
\end{figure*}

The stellar population parameters of our galaxies' analogues are shown in Table \ref{tab:analogue_table}, and compared to observations in Fig. \ref{fig:observation_simulation_ssp}. The $[\alpha/\mathrm{Fe}]$ abundance ratio is not given in the simulations. The ages of our analogues are $20 - 40\%$ younger than those derived from the observations, except for NGC 1351, where the values are consistent.  For total metallicities, both dataset are generally in a good agreement, with NGC 1351 again being an outlier. In our observations, we find that the inner annuli have $\sim 0.12$ dex higher metallicity on average than the outer. On the other hand, for the simulation analogues, the inner annuli are $\sim 0.06$ dex more metal rich on average than the outer annuli, with a larger scatter. For the age gradient, both the observations and simulations have a negligible difference on average, with the simulations having smaller scatter. Considering their weighted averages and uncertainties, both of the datasets sit in a similar region in the $\Delta \mathrm{[M/H] - \Delta \mathrm{Age}}$ diagram. Therefore, assuming that the age and metallicity are the main influence to the PN population, we are confident in using the simulation analogues as an additional tool to interpret our observations. We will use the predicted value of the $M^*$ and the luminosity-specific PN number from PICS to test whether the observed trends are in agreement with the current theoretical understanding.

\begin{table*}[]
    \centering
    \caption{Stellar population age and $\mathrm{[M/H]}$ of the simulated analogue galaxies.}
    \begin{tabular}{c c c c c}
    \hline 
        \multirow{2}{*}{Galaxy analogue} & \multicolumn{2}{c}{Age [Gyr]} & \multicolumn{2}{c}{$\mathrm{[M/H]}$} \\
    \cline{2-5}
        & Inner region & Outer region & Inner region & Outer region\\
    \hline
        NGC 1052   & 8.45 $\pm$ 1.94 & 8.60 $\pm$ 1.69 & 0.09 $\pm$ 0.14 & -0.06 $\pm$ 0.11 \\
        NGC 1351   & 10.07 $\pm$ 1.93 & 9.96 $\pm$ 1.92 & 0.06 $\pm$ 0.04 & -0.01 $\pm$ 0.05\\
        NGC 1380   & 7.65 $\pm$ 1.49 & 8.47 $\pm$ 1.41 & -0.05 $\pm$ 0.23 & -0.12 $\pm$ 0.15 \\
        NGC 1399   & 8.67 $\pm$ 1.40 & 8.48 $\pm$ 1.04 & 0.21 $\pm$ 0.10 & 0.14 $\pm$ 0.10 \\
        NGC 1404   & 9.86 $\pm$ 0.23 & 9.62 $\pm$ 0.31 & 0.14 $\pm$ 0.04 & 0.11 $\pm$ 0.04 \\
    \hline
    \end{tabular}
    \label{tab:analogue_table}
\end{table*}

\begin{figure*}[]
    \centering
    \includegraphics[width=0.85\linewidth]{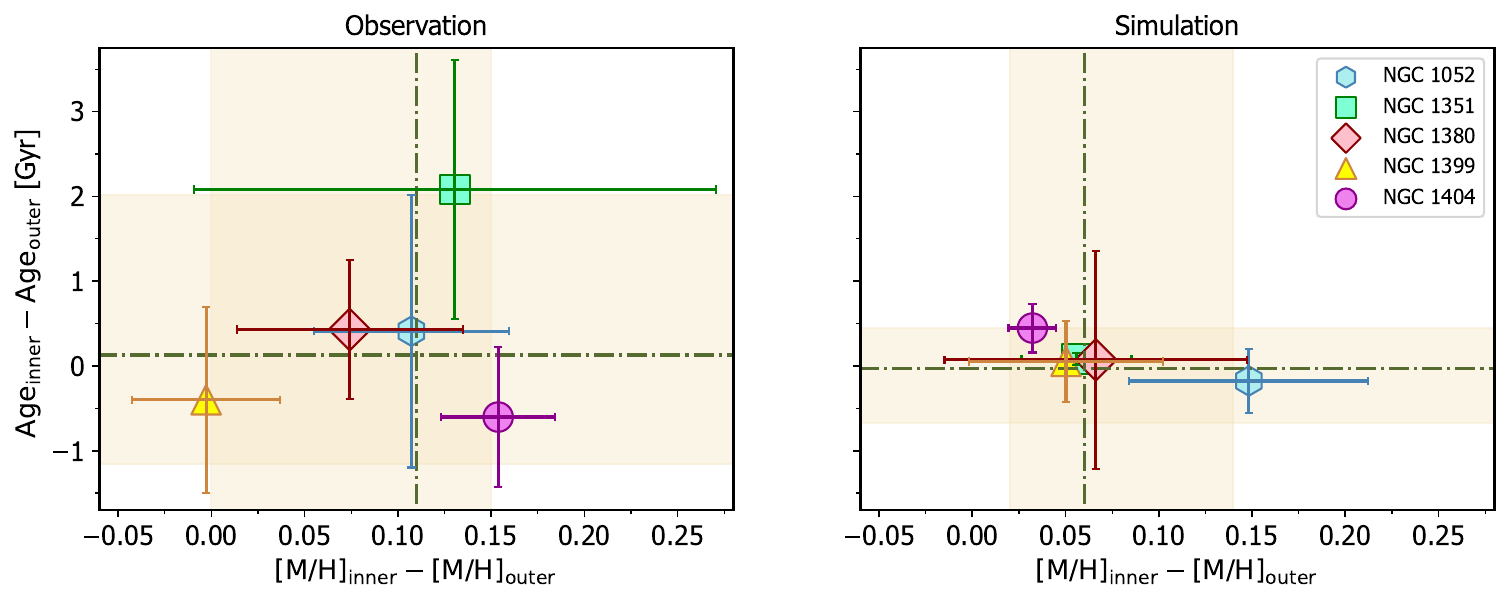}
    \caption{Age and metallicity differences between the inner and outer regions of each observed galaxies (left) and the simulation analogues (right). The Dot-dashed lines and the highlights indicate the weighted average and standard deviation, respectively.}
    \label{fig:differential_difference}
\end{figure*}

\section{Discussion} \label{sec:discussions}

\subsection{Stellar Population Dependence on $M^*$}
\label{sec:Mstar}

To examine the effect of stellar population on the PNLF bright end cut-off, $M^*$, we tested three approaches. First, we compared the observational value of $M^*$ derived using different distance methods. The second approach is from the theoretical perspective, to track the behaviour of $M^*$ in our simulated analogue galaxies. Thirdly, we make differential comparisons of $M^*$ between the inner and outer annulus of the observation and simulation data. Note that this last analysis is differential in nature, and therefore does not require knowledge of the galaxies' of a true distance. 

\subsubsection{$M^*$ from different distance methods}

We derived $M^*$ using three different distances: the PNLF measurements from Paper II, the SBF values from \citet{2009ApJ...694..556B} and \citet{2001ApJ...546..681T}, and the TRGB distances for NGC~1380, 1399, and 1404 \citep{2024ApJ...973...83A}. Currently, there are no TRGB distances to NGC 1351 and NGC 1052;  although TRGB has been measured in the dwarf satellite galaxies NGC~1052-DF2 \citep{2018ApJ...864L..18V} and NGC~1052-DF4 \citep{2020ApJ...895L...4D}, we have opted not to use measurements. 

\begin{figure*}[]
    \centering
    \includegraphics[width=\linewidth]{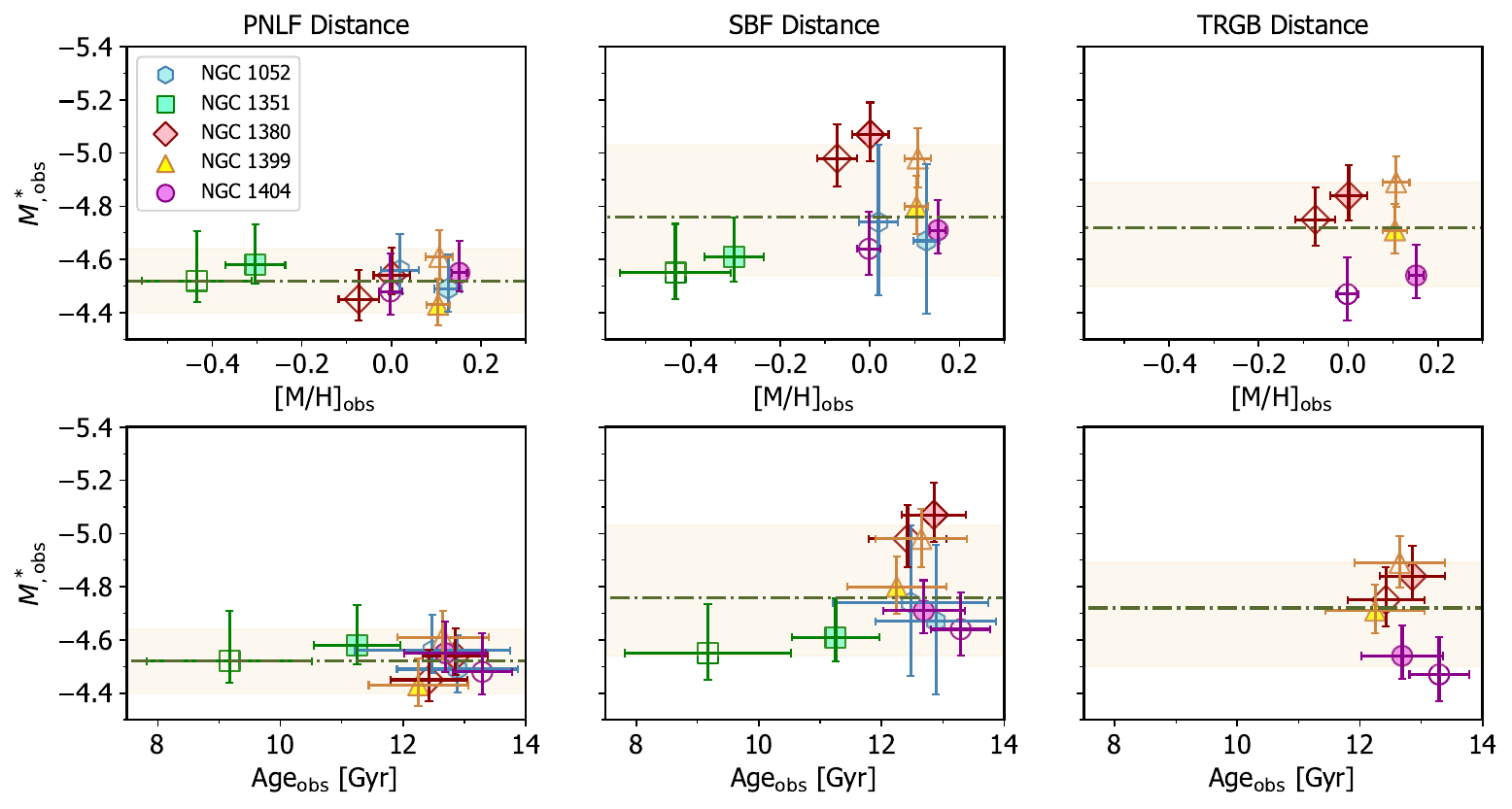}
    \caption{$M^*$ derived using three distance determination methods: PNLF, SBF, and TRGB. The PNLF method assumed $M^* = -4.54$ and act as a reference point for this comparison. Dot-dashed lines indicate the weighted average. Solid and open markers denote the inner and outer annulus, respectively. The highlights indicate the standard deviation.}
    \label{fig:M5007_observation}
    \end{figure*}

A comparison of the relation between $M^*$ and population age and metallicity for the three distance methods is presented in Fig. \ref{fig:M5007_observation}. In Paper I and II, the assumption of $M^* = -4.54$ was used for the distance determination, so it is no surprise that the weighted average of the bright cut-off using PNLF distances is $M^* = -4.52 \pm 0.12$. We use this value as a reference point for the comparison with the $M^*$ values derived from SBF and TRGB distances.

The SBF distances give a weighted average of $M^* = -4.76_{-0.27}^{+0.22}$, which is 0.25 mag brighter than the typically assumed cut-off magnitude. The discrepancy is consistent with the known zero-point offset between the SBF and the PNLF distance scales \citep{2002ApJ...577...31C, 2012Ap&SS.341..151C}. SBF distances tend to have $\sim 0.2$ mag larger distance modulus than the PNLF measurements, possibly because of a combination of minor systematic errors associated with both methods that work in opposite directions \citep{2012Ap&SS.341..151C}.  
The TRGB comparison also yields a PNLF cut-off that is brighter than expectations ($M^* = -4.72_{-0.17}^{+0.22}$), though the nominal value is still within the uncertainties. We note that the $I$-band TRGB measurement typically requires a population with a metallicity $\mathrm{[Fe/H]}\lesssim 0.7$;  otherwise the effect of line-blanketing on the RGB stars is non-negligible, necessitating a metallicity correction \citep{2001ApJ...556..635B, 2002ApJ...577...31C, 2007ApJ...661..815R}.  To minimize this effect, \citet{2024ApJ...973...83A} did select their fields to be in the galaxies' outer regions.  However, some line-blanketing issues may still be present. 

Although solving the discrepancies between the PNLF, SBF, and TRGB distance scales is beyond our scope of this paper, we can still investigate whether metallicity or age have a significant effect on $M^*$. Over the metallicity range $-0.4 \lesssim \mathrm{[M/H]} \lesssim +0.2$ and the age range of 9 to 13.5~Gyr, $M^*$ does not show any trends regardless of the calibration technique, with a side note that we only have one galaxy with a substantial population of stars with ages $<$ 12 Gyr and $\mathrm{[M/H] < -0.2}$.

\subsubsection{$M^*$ from simulated analogue galaxies}

\begin{figure*}[]
    \centering
    \includegraphics[width=0.85\linewidth]{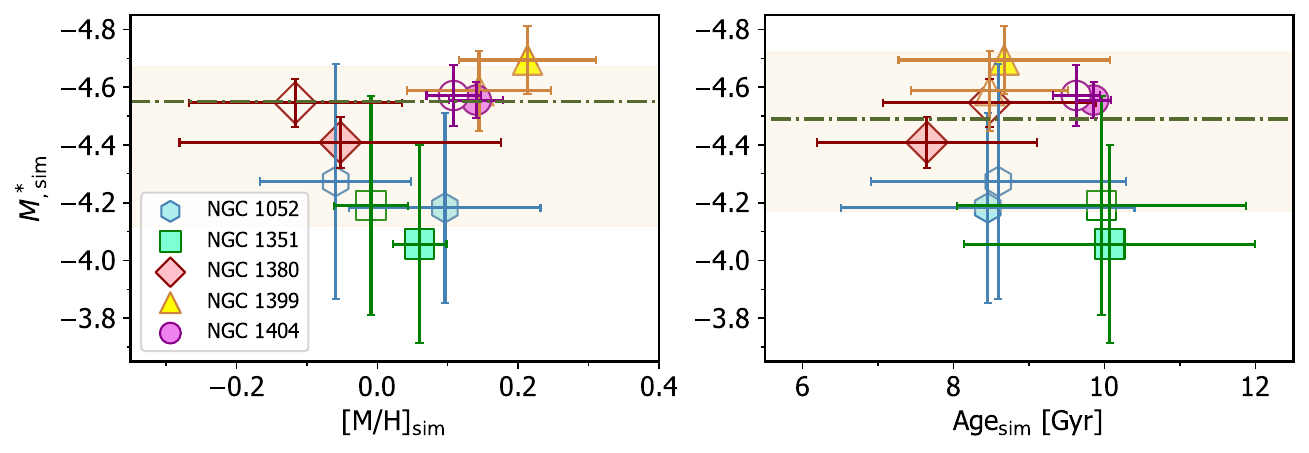}
    \caption{$M^*$ from our simulation analogues as a function of total metallicity (left) and age (right). Dot-dashed lines indicate the weighted average. Solid and open markers represent the inner and outer annuli, respectively. The highlights illustrate the the standard deviation. Note that the $x$- and $y$-axis ranges are different than in Figure  \ref{fig:M5007_observation}.}
    \label{fig:M5007_simulation}
    \end{figure*}

We investigate if a correlation exists between $M^*$, age, and $\mathrm{[M/H]}$ for the simulated galaxies plotted in Fig. \ref{fig:M5007_simulation}. The analogues of NGC 1052 and NGC 1351 have the largest scatter with brightnesses up to $\sim 0.3$ mag fainter than the canonical $M^* = -4.54$ value. Individual analogues with a combination of old ($\gtrsim 10$ Gyr) and metal-poor ($\mathrm{[M/H] \lesssim -0.1}$) tend to produce PN populations with $M^* \sim -4.0$. On the other hand, when the stellar populations are either younger or more metal-rich, the PN populations can exhibit a PNLF cut-off much brighter than $M^* \sim -4.5$. Since some analogues for these two galaxies fall into the earlier category, they should have larger scatter on average. Overall, the simulated galaxies have the weighted average of $M^* = -4.49_{-0.18}^{+0.37}$. We also calculated the weighted average if the two galaxies with the largest scatter are excluded; in that case $M^* = -4.55_{-0.12}^{+0.14}$. Nevertheless, we also do not find any correlation between $M^*$ and age and metallicity for our simulation data.

\subsubsection{$M^*$ from stellar population gradient}

\begin{figure*}[]
    \centering
    \includegraphics[width=0.85\linewidth]{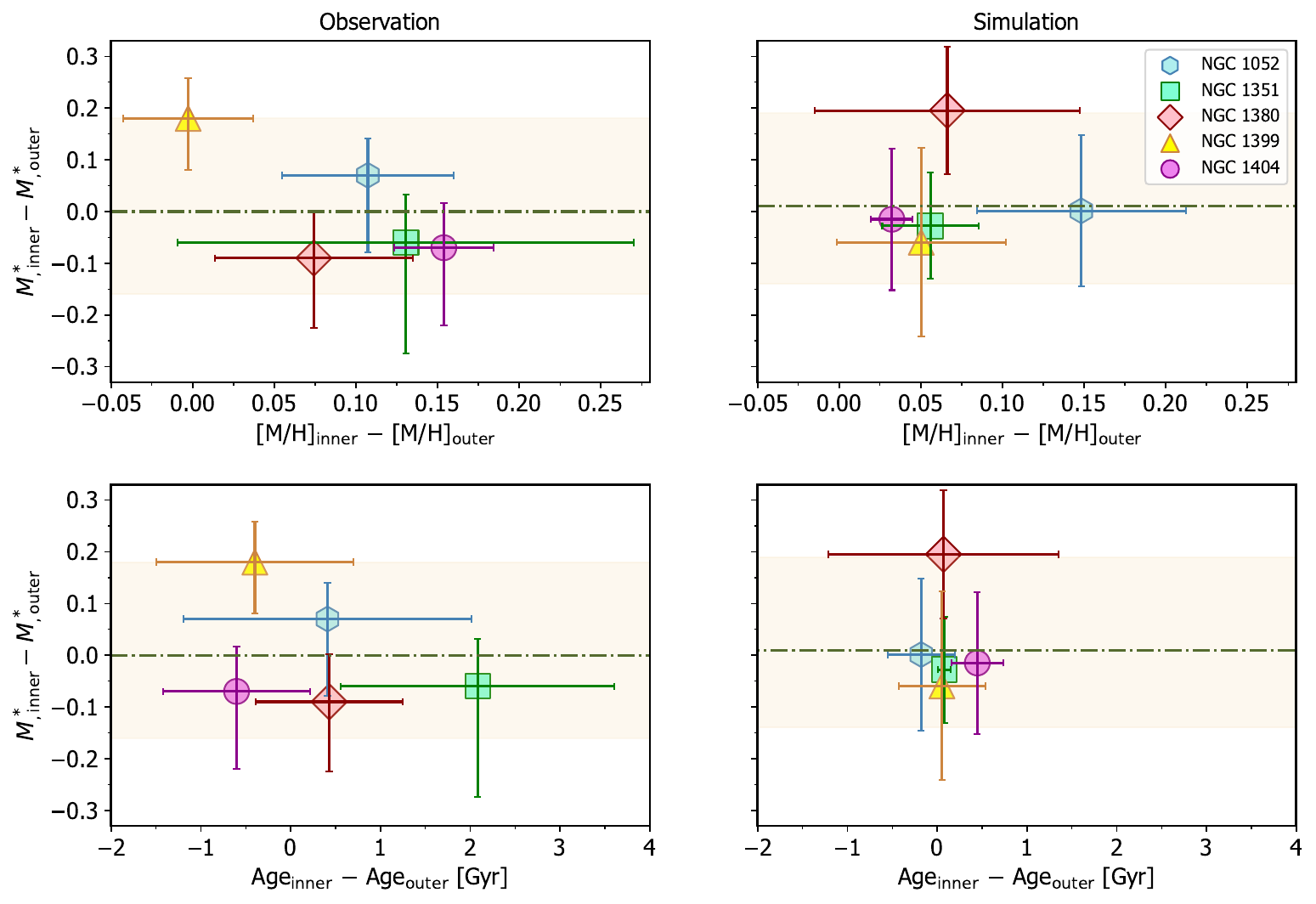}
    \caption{Inner annulus minus outer annulus difference in $M^*$ versus that for 
    metallicity for the observations (left) and simulation analogues (right). Dot-dashed lines indicate the weighted average. The highlights show the standard deviation.}
    \label{fig:M5007_regions}
    \end{figure*}

We now make a differential comparison between the inner and outer annulus of the observations ($\Delta M^*$) and that of the simulation data. This comparison is presented in Fig. \ref{fig:M5007_regions}. In both datasets, we find that $M^*$ is almost identical for the existing age and metallicity gradient within each galaxy. Specifically, we measure a weighted average of $\Delta M^* = 0.00_{-0.16}^{+0.18}$ for the observations and $\Delta M^* = 0.01_{-0.14}^{+0.19}$ for the simulations. In this approach, the stellar population parameters also show a negligible effect on $M^*$. Note that the large $M^*$ uncertainties of the observations stem from the statistical errors on the fits, while the $M^*$ uncertainties associated with the simulations represent the standard deviation from the individual simulated analogue galaxies selected for each observed galaxy.

In their analysis of the PNLF zero-point, \citet{1992ApJ...389...27D, 2010A&A...523A..86S} found that the zero-point discrepancy of the PNLF between Cepheid distance scales should be independent of age and metallicity, which is in agreement with our first approach. In low metallicity systems, \citet{2002ApJ...577...31C} showed that $M^*$ is sensitive to the oxygen abundance and could be up to $\sim 0.2$ mag fainter. For example, in the different populations of M31, \citet{2021A&A...647A.130B} presented the behaviour of $M^*$ as a function of metallicity and showed that between $-0.6 \lesssim \mathrm{[M/H]} \lesssim -0.4$, the PNLF cut-off can be fainter by up to $\sim 1.0$ mag, and reach $M^* \simeq -2.7$ at $\mathrm{[M/H]} \simeq -1.1$.  However, when  $\mathrm{[M/H]} > -0.4$, the PNLF cut-off is expected to stay constant at $M^* \sim -4.5$. A possible fainter $M^*$ due to metallicity has also been observed in M49 \citep{2017A&A...603A.104H}.  Support for this comes from Paper II, where the PNLF cut-off of the galaxy's inner regions was found to be $\sim 0.4$~mag brighter than that derived by \citet{2017A&A...603A.104H} from narrow-band observations of the galaxy's outer envelope and halo. 

Our findings of negligible metallicity dependence in our sample fit into the picture. Since our galaxies have metallicity $\mathrm{[M/H]} > -0.4$, any significant metallicity effect seems unlikely. On the other hand, for populations with super-solar metallicities, \citet{1992ApJ...389...27D} predicted a decrease in $M^*$, which at first glance appears to contradict our results. However, in such systems, there will always be enough stars at solar-type metallicities to produce PNe with $M^* \sim -4.5$.  Hence no shift in $M^*$ should be observable  \citep[see Paper II and][]{2002ApJ...577...31C}.

\subsubsection{Conditions for a constant $M^*$}

Despite our limited sample, we are convinced that the independence of $M^*$ from the stellar population can always be achieved with an appropriate target selection. To elaborate this, we look into the mass-metallicity relationship (MZR\null) of galaxies. The stellar masses of our ETGs are in the range of $10.33 \lesssim \log M_\bigstar/M_{\odot} \lesssim 11.44$ \citep{2019A&A...623A...1I}. Using SDSS, \citet{2006MNRAS.370.1106G} showed that ETGs with $\log M_\bigstar/M_{\odot} \gtrsim 10$ are  likely to have metallicities of $\mathrm{[M/H]} \gtrsim -0.2$. This is also supported by the MZR for $z = 0$ galaxies in the Magneticum simulations \citep{2021ApJ...910...87K, 2025arXiv250401061D}. We did not analyse any star-forming galaxies in this work, but the assumption of a constant PNLF cut-off is also expected to hold for those galaxies. Although their MZR slope is different from that found for quiescent galaxies \citep{2014ApJ...788...72G}, we still expect massive star-forming galaxies to be relatively metal-rich.  Specifically, for star-forming galaxies with stellar mass $\log M_\bigstar/M_{\odot} \gtrsim 10$, we expect  $\mathrm{[M/H]} \gtrsim -0.1$ \citep{2022ApJ...940...32B, 2023ApJ...959...52U,2023ApJ...949...60S, 2024ApJ...960...83S}. The scatter of $\mathrm{[M/H]}$ in these MZRs is typically $\sim0.1$ dex. 

Based on the MZR, metal-poor galaxies with $\mathrm{[M/H]} \lesssim -0.4$ will have low stellar masses, and only a few PNe in the top magnitude of the luminosity function.  Since precision PNLF distances require $\gtrsim 30$~PNe in this magnitude range (Paper I and II), their selection in the context of the Hubble constant measurement should be avoided. If distances were derived from outer halo populations, where the stellar population are metal-poor \citep{2015A&A...579A.135L, 2017A&A...603A.104H, 2020A&A...642A..46H}, a (possibly uncertain) metallicity correction may be necessary \citet{2002ApJ...577...31C}. To minimize this systematic errors, the ideal targets for PNLF-$H_0$ investigations are the inner regions of massive galaxies, where the abundance-based shifts in $M^*$ are negligible.

\subsection{PN numbers at the PNLF cut-off}
\label{sec:apar}

We investigate how the stellar population affects the luminosity-specific PN number, otherwise known as the $\alpha$-parameter, at the PNLF cut-off. Typically, this parameter is defined as $\alpha_{2.5}$, which represents the number of PNe within 2.5 mag of the cut-off per unit bolometric luminosity of the surveyed galaxy, is particularly important as a probe of the underlying stellar population \citep{2006MNRAS.368..877B, 2013A&A...558A..42L, 2017A&A...603A.104H, 2020A&A...642A..46H}. However, as explained in Sect. \ref{sec:PNLF_fits}, we are only interested in PN density at the top 0.5 mag of the luminosity function ($\alpha_{0.5}$). For the simulation data, the $\alpha_{0.5}$ values are obtained directly from the simulation. We discuss its dependence on age, $\mathrm{[M/H]}$ and $[\alpha\mathrm{/Fe]}$. 

\begin{figure*}[]
    \centering
    \includegraphics[width=\linewidth]{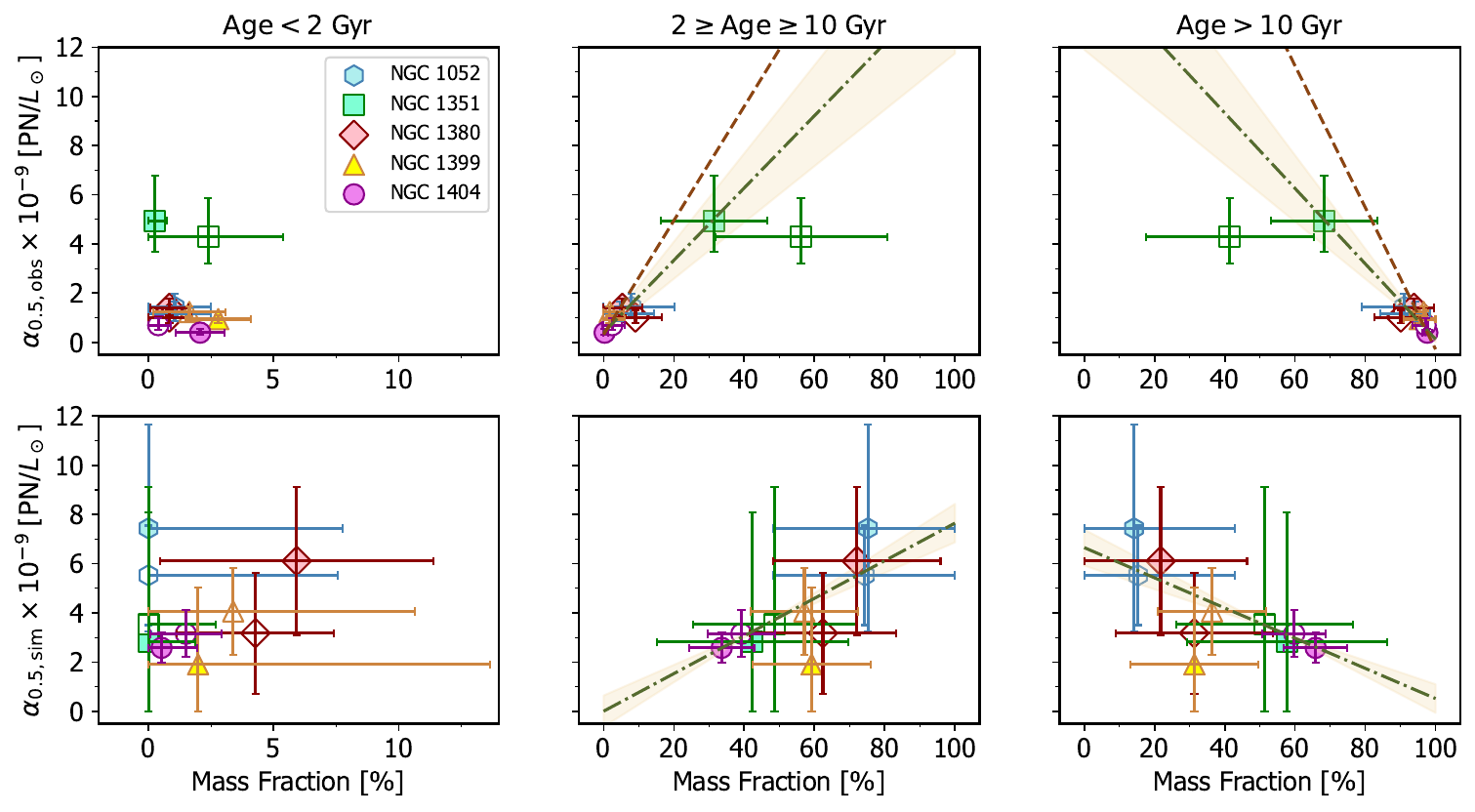}
    \caption{Galaxy $\alpha_{0.5}$ values as a function of the mass fraction of young (Age $< 2$~Gyr), intermediate ($2 \ge \mathrm{Age} \ge 10$~Gyr), and old (Age $> 10$~Gyr) stars.  The upper panels show the observations; the lower panel plot the values from the simulations.  The green dot-dashed lines show the best-fit regressions; fits which exclude the observational data for NGC 1351 are shown via a brown dashed line.}
    \label{fig:alpha_massfraction}
    \end{figure*}

    \begin{figure*}[]
    \centering
    \includegraphics[width=\linewidth]{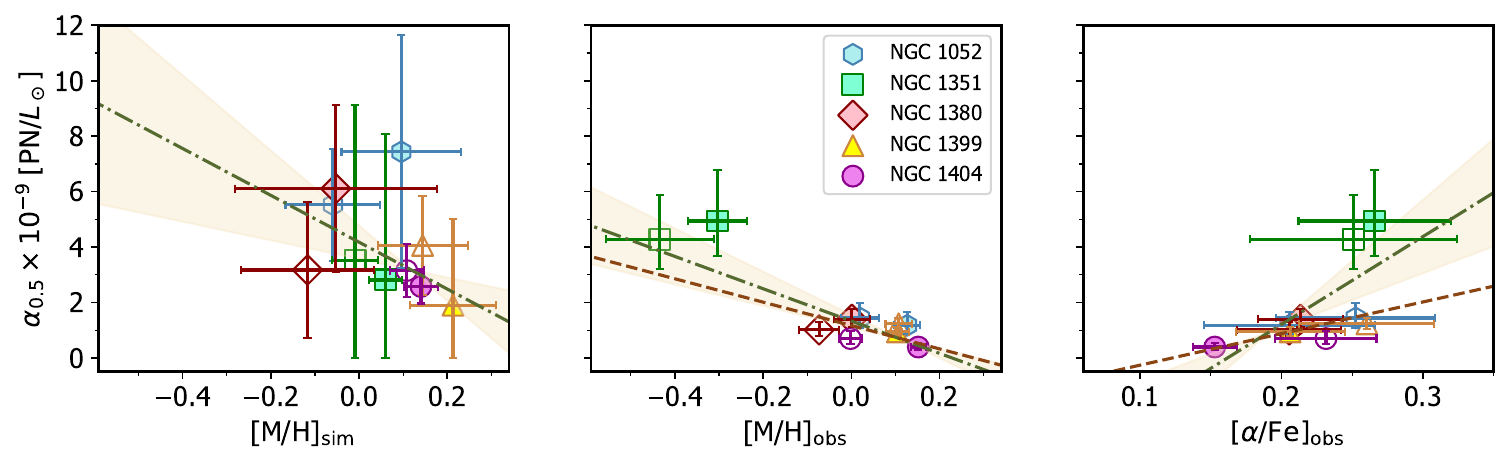}
    \caption{The left and middle panels show
    our $\alpha_{0.5}$ measurements as a function of $[\mathrm{M/H}]$ for the simulations and observations, respectively.  The right panel displays our calculated $[\alpha/\mathrm{Fe}]$ values for the observations.  The green dot-dashed lines show our best fit regressions; fits that exclude NGC 1351 are denoted by the dashed brown line.} 
    \label{fig:alpha_metal}
    \end{figure*}

\subsubsection{Age Dependence} 

We compare the $\alpha_{0.5}$ values of different age groups in Fig. \ref{fig:alpha_massfraction}. As explained in Sect. \ref{sec:SFH}, we defined these age groups based on the core mass predictions for the PNe. Overall, $\alpha_{0.5}$ found from the observations are smaller than those observed in the simulations. However the trends are similar. We tested whether the correlations between $\alpha_{0.5}$ and the mass fractions of different age groups are significant using Pearson correlation statistics with \textit{scipy.stats.pearsonr} \citep{kowalski1972effects}. Then, we fitted a linear function $y = c_1x + c_2$ to the data using orthogonal distance regression with \textit{scipy.odr.odr} \citep{brown1990statistical}. We fitted the data twice for the observation, with and without NGC 1351. The fit parameter results are presented in Table \ref{tab:slope_ages}.

\begin{table*}[]
    \centering
    \caption{Linear fit parameters between $\alpha_{0.5}$ and mass fraction of different ages.}
    \begin{tabular}{c c c c c c c}
    \hline
        \multirow{2}{*}{Data} & \multicolumn{3}{c}{Intermediate Ages} & \multicolumn{3}{c}{Old Ages} \\
    \cline{2-7}
        & $p$-value & $c_1$ & $c_2$ & $p$-value & $c_1$ & $c_2$\\
    \hline
        Observations with NGC 1351 & 0.001 & 0.14 $\pm$ 0.03 & 0.38 $\pm$ 0.10 & 0.001 &-0.15 $\pm$ 0.04 & 14.67 $\pm$ 3.38 \\
        Observations without NGC1351 & 0.11 & 0.22 $\pm$ 0.07 & 0.30 $\pm$ 0.12 & 0.11 & -0.27 $\pm$ 0.10 & 27.09 $\pm$ 10.00\\
        Simulations                  & 0.01 & 0.08 $\pm$ 0.02 & 0.00 $\pm$ 0.72 & 0.02 & -0.06 $\pm$ 0.01 & 6.65 $\pm$ 0.73 \\
    \hline
    \end{tabular}
    \label{tab:slope_ages}
\end{table*}

Both observation and simulations contain $\lesssim 5 \%$ mass fractions for stars younger than 2 Gyr. In neither, do we see any correlation between the young population and $\alpha_{0.5}$. The $p$-values for the young populations are 0.76 and 0.84 for the observations and simulations, respectively; we skipped the fitting process due to the non-correlation. For the intermediate age population, we found a positive linear correlation for both the observations and the simulations, though the correlation for the observations weakens when NGC\,1351 is excluded. Although the slope derived from the observation is steeper than that predicted from the models, the main indication is that more bright PNe will be formed when the mass fraction of the intermediate age population is larger. This is in agreement with \citet{2019ApJ...887...65V, 2025A&A...699A.371V}, who predicted most PNe observed at the bright-end of the PNLF have main sequence lifetimes between 2 and 10 Gyr. Moreover, we found a negative correlation when comparing $\alpha_{0.5}$ to the $> 10$~Gyr stellar population.

These findings imply a relationship between PN core-mass and $\alpha_{0.5}$. The PN formation rate per luminosity unit is rather insensitive of age \citep{1986ASSL..122..195R, 2006MNRAS.368..877B}. However, the visibility timescale of a PN at its \oiii-bright phase depends strongly on the mass of its central star, and this mass, in turn depends on the progenitor's initial mass and therefore lifetime. As a result, PN visibility depends critically on the inverse of population age, and this largely defines $\alpha_{0.5}$. For example, when the population is very young ($<2$ Gyr), the PN \oiii-bright phase timescale will be very short, and theoretically should make the $\alpha_{0.5}$ smaller; this trend is not visible in our ETGs, possibly because all of our samples exhibit similar mass fraction of the very young populations. The positive trend for the intermediate ages show that an older population in this age range create PNe with lower core mass, which evolve slower, and therefore stay visible longer; giving higher $\alpha_{0.5}$ values. For the old populations ($>10$ Gyr), a different condition occur. While the lifetimes of these stars are the longest, a core-mass of $\lesssim 0.53 \: \mathrm{M}_\odot$ simply do not produce enough luminosity to produce $\gtrsim 400 L_{\odot}$ of \oiii~emission \citep{2019ApJ...887...65V}. Hence, the $\alpha$-parameter will decrease if the fraction of old population increases.

\subsubsection{$\mathrm{[M/H]}$ and $[\alpha\mathrm{/Fe]}$ Dependence}

We also analysed the trend of $\alpha_{0.5}$ with respect to metallicity. For the observations, we looked into correlation with both $\mathrm{[M/H]}$ and $[\alpha\mathrm{/Fe]}$; for the simulations, we only examined $\mathrm{[M/H]}$, since no information on the $\alpha$-process to iron ratio was available. We tested the correlations and fitted the data similarly with the age fractions.  The relations are presented in Fig. \ref{fig:alpha_metal} and the fitted parameters are listed in Table \ref{tab:slope_metal}. We find an anti-correlation between $\alpha_{0.5}$ and $\mathrm{[M/H]}$ in both datasets, but a positive correlation between $\alpha_{0.5}$ and $[\alpha\mathrm{/Fe]}$.  This is expected, because in ETGs, $\mathrm{[M/H]}$ and $[\alpha\mathrm{/Fe]}$ are anti-correlated \citep{2015A&A...582A..46W}. However, we should note that the correlation with $\mathrm{[M/H]}$ relies on NGC\,1351:  if that galaxy is excluded, the result is no longer significant.

Previously, \citet{2005ApJ...629..499C} investigated the metallicity dependence of $\alpha_{0.5}$ using the $\mathrm{[MgFe]}$ line indices and found no strong trend. Similarly, in their MUSE study of edge-on galaxies in Fornax, \citet{2021A&A...652A.109G} measured the $\alpha_{2.5}$-$\mathrm{[M/H]}$ relationship via two different methods: full spectral fitting and single-age population fits focused on specific line indices. They also found an anti-correlation of $\alpha_{2.5}$ with $\mathrm{[M/H]}$ but no trend versus the line indices. They speculated that the discrepancy might be caused by the variation in $[\alpha\mathrm{/Fe]}$. We speculate that the enrichment of alpha elements, which also includes oxygen, plays a role in the production efficiency of PNe at the PNLF cut-off. Assuming our derived trend which includes NGC\,1351 is accurate, one could have a factor of $\sim 3-5$ times more PNe at the top 0.5 mag of the PNLF for a system with $\mathrm{[M/H]} = -1.0$. This prediction aligns to the findings of high $\alpha$-parameters in outer haloes of ETGs, where the metallicity is expected to be low \citep{2013A&A...558A..42L, 2015A&A...579A.135L, 2017A&A...603A.104H, 2020A&A...642A..46H}. In the metallicity range where the $M^*$ can be assumed as constant, i.e., the central $R_e$ of ETGs, $\alpha_{0.5}$ should be relatively constant and only vary by a factor of $\lesssim 2$. 

\begin{table*}[]
    \centering
    \caption{Linear fit parameters of  $\alpha_{0.5}$ with $\mathrm{[M/H]}$ and $[\alpha\mathrm{/Fe]}$.}
    \begin{tabular}{c c c c c c c}
    \hline
        \multirow{2}{*}{Data} & \multicolumn{3}{c}{$\mathrm{[M/H]}$} & \multicolumn{3}{c}{$[\alpha\mathrm{/Fe]}$} \\
    \cline{2-7}
        & $p$-value & $c_1$ & $c_2$ & $p$-value & $c_1$ & $c_2$\\
    \hline
        Observations with NGC 1351   & 0.001 &-3.97 $\pm$ 1.47 & 1.17 $\pm$ 0.21 & 0.05 & 27.13 $\pm$ 12.59 & -4.03 $\pm$ 2.28 \\
        Observations without NGC1351 & 0.48 & -2.59 $\pm$ 1.34 & 0.99 $\pm$ 0.20 & 0.06 & 9.56 $\pm$ 3.77 & -1.00 $\pm$ 0.67\\
        Simulations                  & 0.33 & -8.45 $\pm$ 5.10 & 4.18 $\pm$ 0.60 & - & - & - \\
    \hline
    \end{tabular}
    \label{tab:slope_metal}
\end{table*}

\subsection{PN progenitors at the PNLF cut-off}

\begin{figure*}[]
    \centering
     \includegraphics[width=0.85\linewidth]{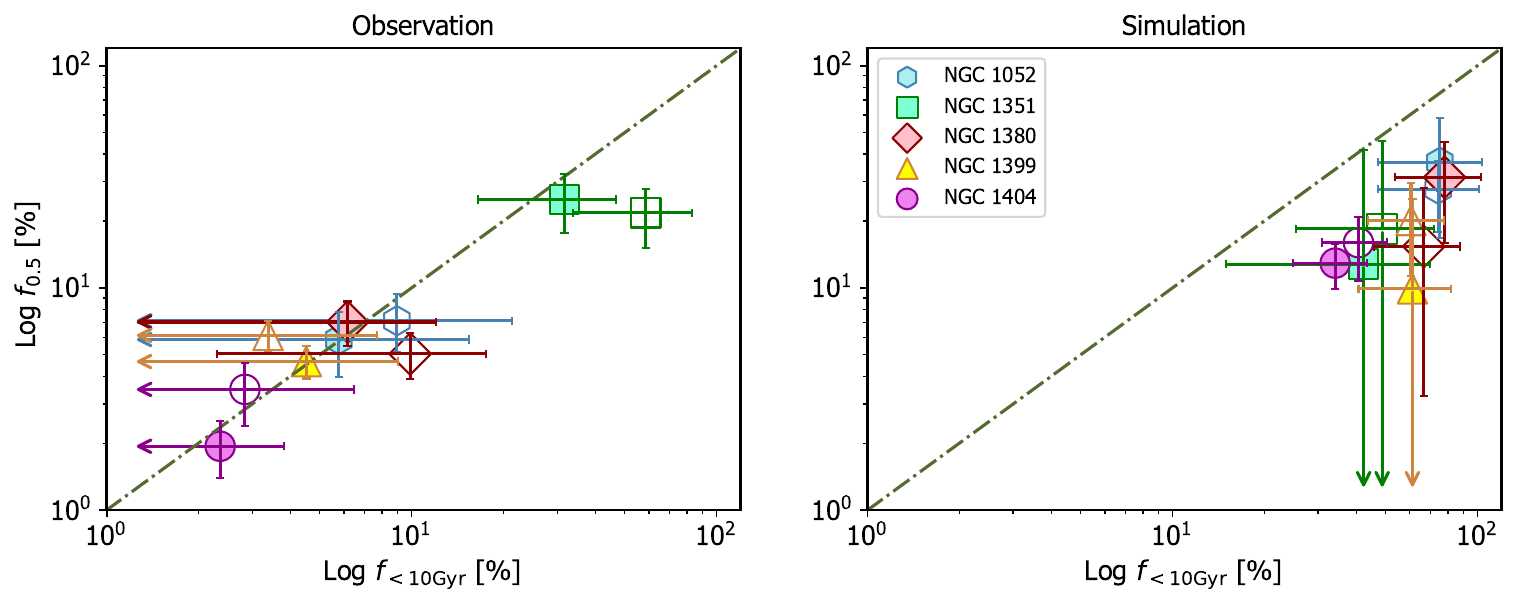}
    \caption{The production efficiency for PNe within 0.5~mag of $M^*$ versus the mass fraction of stars with ages less than 10~Gyr for the five galaxies in our survey.  The left panel shows the observations data; the right panel shows the predictions from the simulated galaxies.  The arrows indicate uncertainties that reach values of less than $1\%$.}
    \label{fig:mass_fraction_alpha05}
    \end{figure*}

From the $\alpha_{0.5}$ values, we can also estimate the fraction of stars that turned into the PNe at the top 0.5 mag of the PNLF ($f_{0.5}$).  Based on the fuel-consumption theorem \citep{1986ASSL..122..195R, 2006MNRAS.368..877B}:
\begin{equation}
    \alpha_{0.5} = B \times t \times f_{0.5}
    \label{eq:alpha}
\end{equation}
\noindent
where $B$ is the bolometric luminosity-specific stellar evolutionary flux and $t$ is the visibility lifetime of the PN-phase when $M^*$ is at its maximum. For all PN-forming stellar populations $B \simeq 2 \times 10^{-11} \: \mathrm{stars} \: \mathrm{yr}^{-1} \: L_\odot^{-1}$, with a minor dependency on age and IMF\null.  Based on post-AGB evolution models \citep{2016A&A...588A..25M}, cores bright enough to excite a PN within 0.5~mag of $M^*$
($\sim 0.58$ to $0.62 M_{\odot}$), only have visibility timescales of 500 to 1500 years \citep[e.g.,][]{2007A&A...473..467S, 2018NatAs...2..580G}. If we adopt 1000 years for this \oiii-bright phase, then we can compare $f_{0.5}$ to the mass fraction of a galaxy's stars that are younger than 10 Gyr ($f_{< \:10 \mathrm{Gyr}}$). We combine the young and intermediate-age stellar population because both can theoretically produce PNe in the top 0.5 mag of the luminosity function \citep{2019ApJ...887...65V, 2025A&A...699A.371V}. The comparison is shown in Fig. \ref{fig:mass_fraction_alpha05}. 

Despite the large uncertainties from the stellar population synthesis, our assumptions produce an almost one-to-one relation between the fraction of stars derived from the fuel-consumption theory and that inferred from the galaxies' young and intermediate-age stellar populations.  This seems to imply that the ETGs have enough young and intermediate-mass stars to produce the PNe seen in the top 0.5~mag of the PNLF\null. However, there is more to the story. We assume PN formation efficiencies at the top 0.5~mag using
\begin{equation}
    \eta_{{\rm PN}, 0.5} = \frac{f_{0.5}}{f_{< \:10 \mathrm{Gyr}}} \times 100\%
\end{equation} 
Our observations, imply a median efficiency of $90 \pm 40\%$; such a high efficiency and rather unlikely.  For example, when we apply our simple assumptions to the simulations, we predict that only $35 \pm 10\%$ of the stars turned into the \oiii-bright PNe. Moreover, we argue that even this number is already an overestimation for two main reasons: (1) the simulations currently predict more PNe than the observations, and (2) the core masses obtained from the simulations have a median mass of $0.55 \:\mathrm{M}_\odot$; such a low mass implies a visibility timescale that is almost an order of magnitude longer than that assumed.  Thus, through Eq. \ref{eq:alpha}, $f_{0.5}$ and therefore the actual PN production efficiency must be much less than $35\%$.  

There are several arguments to explain the discrepancy between our calculated value of 90\% efficiency, and the expectations that the number should be of the order of $\sim 10\%$.   Firstly, if the fraction of a given population is small (say, less than 5\%), then the uncertainty in the measurement is generally quite large, typically between 50\% and 370\%.  The detection of such a small population is very challenging and the results propagate directly into the efficiency predictions. At the extreme, it could reduce the efficiency down to $\sim20\%$. 

Secondly, the assumed visibility timescale of 1000 years is based on post-AGB models for stellar population age of $\sim 1 $ Gyr \citep{2018NatAs...2..580G}. If older populations also produce $M^*$ PNe, then the mean visibility timescale of the observed PNe will be longer.  In fact, a small difference in the age of the assumed PN population could easily change the timescale by a factor of two, and this error would immediately propagate into calculated efficiency.  Similarly, the assumed age population also effects $B$\null.  Although the luminosity specific stellar evolution flux, is relatively independent of stellar population, changes in age and IMF can shift its value by $10 -20\%$.  Again, this change propagates directly into the efficiency estimate.

Regardless on the true value of $\eta_{{\rm PN}, 0.5}$, the linear relationships between $f_{0.5}$ and $f_{< \:10 \mathrm{Gyr}}$ in both observations and simulations has an important implication on the PN formation mechanism in old stellar systems. Previously, it was thought that a population of $\sim 1$ Gyr was needed to form $M^*$ PNe.  Since the fraction of such stars in ETGs is extremely low, other evolutionary channels were invoked.  These included linking PNe to products of  binary evolution, such blue stragglers stars \citep{2005ApJ...629..499C} or accreting white dwarfs \citep{2006ApJ...640..966S, 2023MNRAS.521.1808S}. This might be no longer necessary.  The latest generation of post-AGB models \citep{2016A&A...588A..25M}, when coupled with a newer IFMR allows allows stellar populations up to 10 Gyr in age to  form $M^*$ PNe \citep{2025A&A...699A.371V, 2025ApJ...983..129J}.  A direct consequence of this change is the \oiii-bright phase of PNe is now predicted to be longer, and as illustrated above, will significantly influence the efficiency of PN production in ETGs. This highlights the importance of the IFMR for the physics of the $M^*$ PNe and the PNLF as standard candles.

\section{Conclusions and Outlook} \label{sec:conclusion}

We analysed five early-type galaxies (ETGs) using MUSE archival data. In each galaxy, we defined two annuli, identified the regions' planetary nebulae, and measured PNLFs' cut-off magnitudes ($M^*$) and luminosity-specific density of PNe in the top 0.5 mag of the luminosity function ($\alpha_{0.5}$). Then, for each annulus, we derived the underlying stellar population's age, $\mathrm{[M/H]}$, and $[\alpha\mathrm{/Fe]}$ using \texttt{pPXF}\null. Additionally, we applied the PICS model \citep{2025A&A...699A.371V} to simulated analogue galaxies from Magneticum Pathfinder to interpret our findings. We studied the correlation between the PNLF quantities ($M^*$ and $\alpha_{0.5}$) and the underlying stellar parameters (age, $\mathrm{[M/H]}$, and $[\alpha\mathrm{/Fe]}$), with our main goal being to prepare for our PNLF-$H_0$ effort by testing how the PNLF cut-off changes with stellar population.  We also discussed the implications of our analysis regarding the question of the progenitors of the most luminous PNe in very old stellar systems. Our conclusions are the following:

\begin{enumerate}
    \item Four of our ETGs have stellar populations with $12\lesssim \mathrm{Age} \lesssim13.5$ Gyr and $-0.1\lesssim \mathrm{[M/H]} \lesssim+0.2$; the fifth galaxy, NGC\,1351, has an average age of $\sim 10$ Gyr and an average metallicity of $\mathrm{[M/H]} \sim -0.3$. We find no significant dependence of $M^*$ with population age or $\mathrm{[M/H]}$.  While there are zero-point discrepancies on $M^*$ when assuming different PNLF distance calibrators, these are unlikely to be related to the galaxies' stellar populations. The constancy of $M^*$ is supported by the mass-metallicity relationship (MZR) in galaxies; more massive galaxies will have higher metallicities, and therefore $M^*$ will be invariant. Based on this, massive ETGs, and in particular their inner regions, will have metallicities that do not have any effect on $M^*$, whilst producing enough PNe to ensure the bright-end of the PNLF is well populated.  Such conditions are ideal for PNLF distance determination.  

    \item The luminosity-specific number of PNe in the top 0.5 magnitude of the PNLF ($\alpha_{0.5}$) is crucial for understanding the formation efficiency of \oiii-bright PNe. We find a strong positive correlation of $\alpha_{0.5}$ with the fraction of the intermediate age ($2 < \mathrm{Age} <10$ Gyr) stars in our galaxies. Conversely, $\alpha_{0.5}$ decreases when the population of older stars ($t > 10$~Gyr) dominates, and is insensitive to stellar populations younger than 2 Gyr. Moreover, we find that $\alpha_{0.5}$ does not depend on metallicity when $\mathrm{[M/H]} > -0.1$, but increases below this value. We also see a slight increasing trend of $\alpha_{0.5}$ with $[\alpha\mathrm{/Fe]}$. 

    \item Using the fuel-consumption theorem \citep{1986ASSL..122..195R, 2006MNRAS.368..877B} and population synthesis models, we find a significant linear relation between the fraction of stars turned into PNe at the top 0.5 mag of the luminosity function ($f_{0.5}$) and the mass fraction of the stellar population younger than 10 Gyr ($f_{< \:10 \mathrm{Gyr}}$). Based on our sample, we show that, with at least $2\%$ of $f_{< \:10 \mathrm{Gyr}}$, old stellar systems can in principle produce $M^*$ PNe if the stars have sufficiently long lifetimes. This is supported by recent findings of \citet{2025A&A...699A.371V} and \citet{2025ApJ...983..129J}, in that variations of the initial-final mass relation (IFMR) allow stellar systems as old as 10 Gyr to produce PNe that occupy the brightest 0.5~mag of the PNLF; such old progenitors will have longer evolutionary lifetimes and subsequently longer visibility timescales.      

    \item The PICS models confirm the trends seen in our observations. The models show no correlation between $M^*$ and the stellar population age and $\mathrm{[M/H]}$ in the simulated sample. Despite the simulations having systematically younger average stellar ages, the general trends between $\alpha_{0.5}$ and mass fractions of different ages follow those of the observations. There is also no significant correlation between $\alpha_{0.5}$ and $\mathrm{[M/H]}$ at $\mathrm{[M/H]} \gtrsim -0.1$, similar to the observations. The significant linear relation between $f_{0.5}$ and $f_{< \:10 \mathrm{Gyr}}$ also exists in the simulation data, despite a zero-point difference in the relationship.  

\end{enumerate}

To better constrain our knowledge of the formation efficiency of the brightest PNe, more precise measurements of the minority stellar mass fractions of a galaxy are necessary. Deep learning based stellar population studies might improve the precision over stellar population fitting tools, such as \texttt{pPXF} \citep{2024MNRAS.530.4260W}. Furthermore, \citet{2024MNRAS.528.2790Z} speculate that non-parametric methods for recovering SFHs like \texttt{pPXF} might have hidden intrinsic biases that still requires further investigations. Updated time-dependent PN hydrodynamical simulations, which consider different IFMRs, can also provide a better estimate of the lifetime of the \oiii-bright phase for PNe with lower core masses. With improved distances of Milky Way (MW) PNe, it is also possible to probe the local PNLF \citep{2023arXiv231117820C}. Spatially resolved spectroscopic follow-up of the PNe at the MW PNLF cut-off can help us understand how they are formed in greater details than studies or extragalactic PNe (Soemitro et al., in prep). Finding low-mass central stars at the PNLF cut-off would be particularly relevant for explaining PNe in ETGs. MW PNe can also reveal in how important binary interactions are for the formation process of $M^*$ PNe in early-type systems.

\begin{acknowledgements}
      Based on data obtained from the ESO Science Archive Facility with DOIs \url{https://doi.org/10.18727/archive/41} and \url{https://doi.org/10.18727/archive/42}. This research has made use of the NASA/IPAC Extragalactic Database, which is funded by the National Aeronautics and Space Administration and operated by the California Institute of Technology. This work made use of Astropy:\footnote{http://www.astropy.org} a community-developed core Python package and an ecosystem of tools and resources for astronomy \citep{astropy:2013, astropy:2018, astropy:2022}. AS and MMR acknowledge support from DFG under grants RO 2213/40-1, RO 2213/41-1, RO 2213/42-1, RO 2213/43-1. LMV acknowledges support by the German Academic Scholarship Foundation (Studienstiftung des deutschen Volkes) and the Marianne-Plehn-Program of the Elite Network of Bavaria. The Institute for Gravitation and the Cosmos is supported by the Eberly College of Science and the Office of the Senior Vice President for Research at the Pennsylvania State University. This work was partially supported by the NSF through grant AST2206090. GSC acknowledge support from DFG under grant CO 3160/1-1.
\end{acknowledgements}

\bibliographystyle{aa}
\bibliography{muse-pnlf-3}

\clearpage
\onecolumn

\begin{appendix}

\section{SSP Library Comparison}
\label{appendix:SSP_comparison}

Stellar population parameters obtained from spectral fitting procedures are model dependent, as mentioned in Sect. \ref{subsec:popmodels}. To explore the variation that might occur, we compared the fiducial results obtained with the sMILES SSP models \citep{2023MNRAS.523.3450K} with those calculated using MILES SSPs  \citep{2015MNRAS.449.1177V} and the semi-empirical library of \citet{2009MNRAS.398L..44W}, to be referred as W09. The parameters for each SSP are presented in Table \ref{tab:param_SSPs}. 

\begin{table}[h!]
    \centering
    \caption{Parameter grid of the SSP models.} \label{tab:param_SSPs}
    \begin{tabular}{c c c c}
    \hline \hline
        Parameters & sMILES & MILES & W09 \\
    \hline
            & 0.03, 0.04, 0.05, 0.06, 0.07,& 0.03, 0.04, 0.05, 0.06, 0.07, &  \\
            & 0.08, 0.09, 0.10, 0.15, 0.20,& 0.08, 0.09, 0.10, 0.15, 0.20, &  \\
            & 0.25, 0.30, 0.35, 0.40, 0.45,& 0.25, 0.30, 0.35, 0.40, 0.45, &  \\
            & 0.50, 0.60, 0.70, 0.80, 0.90,& 0.50, 0.60, 0.70, 0.80, 0.90, &  \\
            & 1.00, 1.25, 1.50, 1.75, 2.00,& 1.00, 1.25, 1.50, 1.75, 2.00, & 2.00, 3.00, 4.00, 5.00, 6.00,\\
            Ages $[\mathrm{Gyr}]$ & 2.25, 2.50, 2.75, 3.00, 3.25,& 2.25, 2.50, 2.75, 3.00, 3.25, & 7.00, 8.00, 9.00, 10.0, 11.0,\\
            & 3.50, 3.75, 4.00, 4.50, 5.00,& 3.50, 3.75, 4.00, 4.50, 5.00, & 12.0, 13.0 \\
            & 5.50, 6.00, 6.50, 7.00, 7.50, & 5.50, 6.00, 6.50, 7.00, 7.50, & \\
            & 8.00, 8.50, 9.00, 9.50, 10.0, & 8.00, 8.50, 9.00, 9.50, 10.0, &  \\
            & 10.5, 11.0, 11.5, 12.0, 12.5, & 10.5, 11.0, 11.5, 12.0, 12.5, & \\
            & 13.0, 13.5, 14.0 & 13.0, 13.5, 14.0 &\\
    \hline

            & -1.79, -1.49, -1.26, -0.96, & -1.79, -1.49, -1.26, -0.96, & \\
     $[\mathrm{M/H}]$ & -0.66, -0.35, -0.25,  0.06, & -0.66, -0.35, -0.25,  0.06, & -- \\
            & 0.15, 0.26 & 0.15, 0.26 & \\
    \hline

    $[\mathrm{Fe/H}]$& -- & -- & -0.50 , -0.25,  0.00  ,  0.20 \\

    \hline
    
    $[\alpha/\mathrm{Fe}]$ & -0.20, 0.00 , 0.20, 0.40, 0.60 & 0.00, 0.40 & -0.20, 0.00, 0.20, 0.40 \\

    \hline
    \end{tabular}
\end{table}

For every SSP library, we followed the same fitting routine described in Sect. \ref{subsec:fitting}. We compared the age, the mass-weighted age fractions, $[\mathrm{M/H}]$, and $[\alpha/\mathrm{Fe}]$. This comparison should give an impression regarding the model dependency of our measurements. Since the W09 abundances are given in $[\mathrm{Fe/H}]$ and $[\alpha/\mathrm{Fe}]$, we calculated the total metallicity using 

\begin{equation}
    [\mathrm{M/H}] = [\mathrm{Fe/H}] + 0.75 \times [\alpha\mathrm{/Fe}]
\end{equation}

\noindent
We discuss the comparison for each parameter in the sub-sections below. 

\subsection{Age}
 
\begin{figure}
    \centering
    \includegraphics[width=0.8\linewidth]{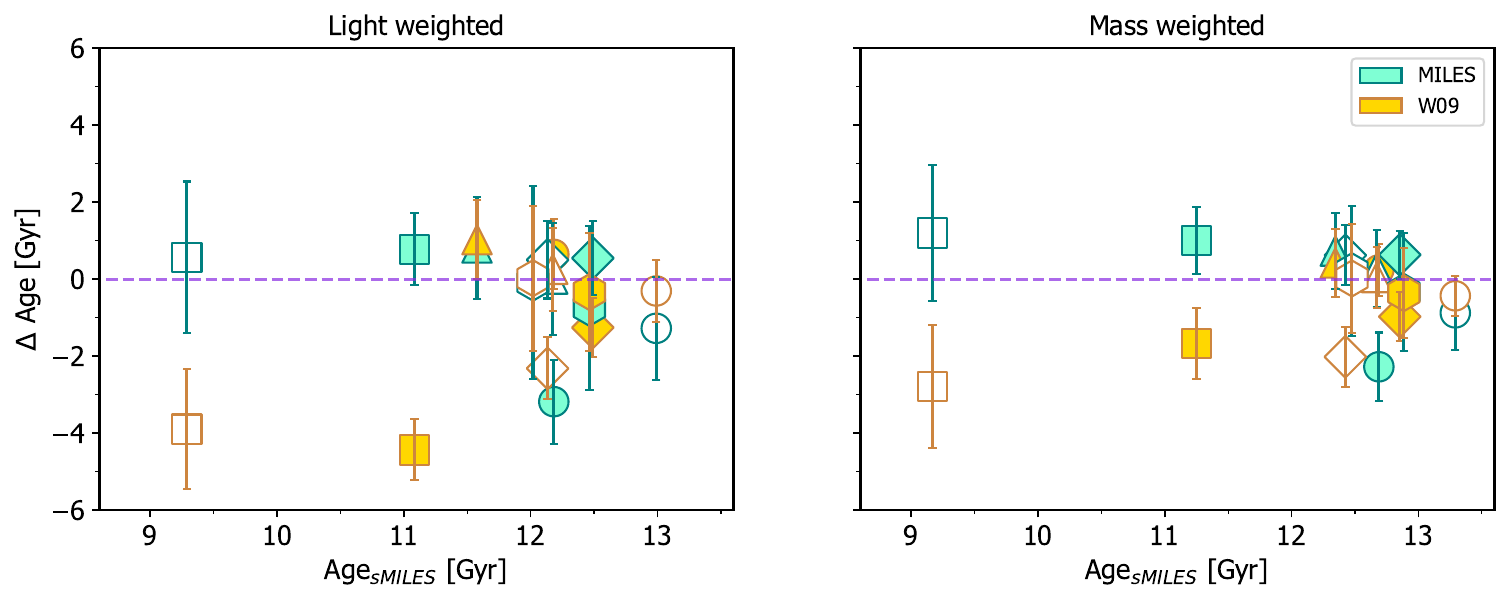}
    \caption{Differences of average weighted ages between the results of the sMILES models and those from comparison models. MILES predictions are indicated in green and WO9 results with yellow. The galaxy markers are: NGC 1052 (hexagon), NGC 1351 (square), NGC 1380 (diamond), NGC 1399 (triangle), and NGC 1404 (circle). Filled and empty markers indicate the inner and outer region, respectively.}
    \label{fig:age_comparison}
\end{figure}

\begin{figure}
    \centering
    \includegraphics[width=0.9\linewidth]{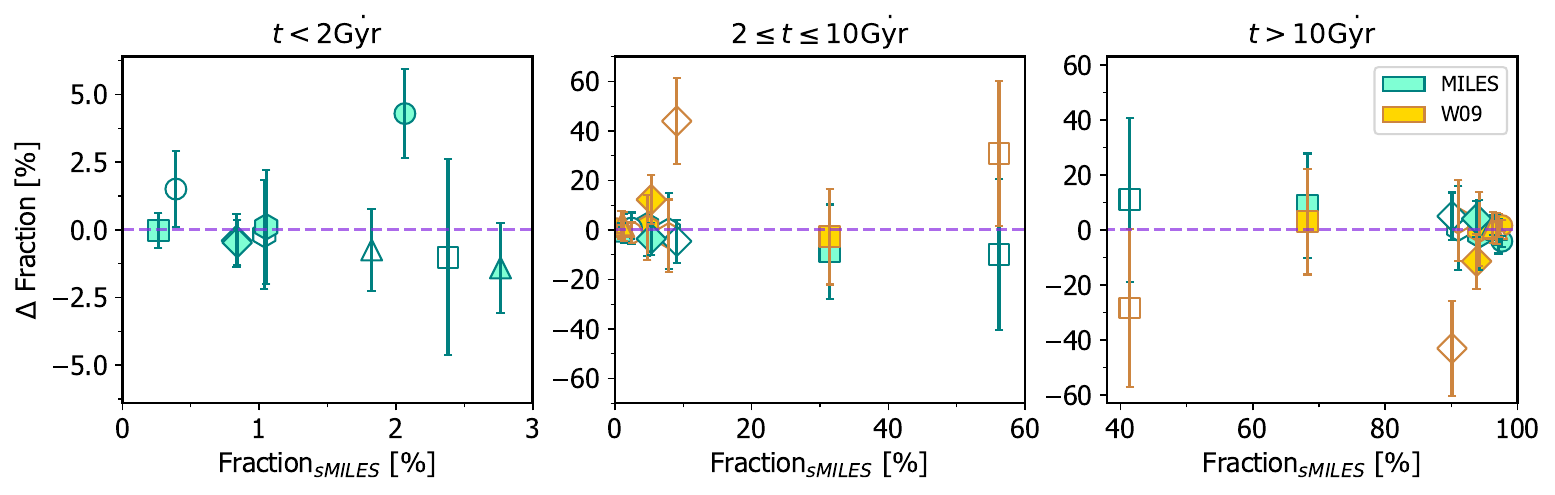}
    \caption{Differences of mass-weighted age fractions between the results of the sMILES models and those of comparison models. The marker descriptions are the same as those in Fig. \ref{fig:age_comparison}.}
    \label{fig:sfh_comparison}
\end{figure}

The ages produced by the three SSP models are in general agreement considering the error bars; this can be seen in Fig. \ref{fig:age_comparison}. However, some exceptions exist. In the comparison with the MILES library, the largest differences occur for the inner region of NGC\,1404, where the MILES models produced an age that was $\sim$ 3 Gyr younger than our fiducial sMILES age. Stemming from how the templates are weighted, the MILES models also recover $\sim$ 5\% more stars younger than 2 Gyr, compared to the sMILES models (see Fig. \ref{fig:sfh_comparison}). For the comparison with W09 models, the largest differences are found in the outer region of NGC\,1380 and both regions of NGC\,1351, where W09 produced younger ages by up to 4 Gyr. In the outer regions of NGC\,1351 and NGC\,1380, these discrepancies might be explained by the larger ($\sim$ 40\%) fraction of intermediate-age stars that the W09 models assigns; this come at the expense of older population, which is 30\% to 40\% smaller.  For the inner region of NGC 1351, there are no notable differences in our age fraction plots. However, this is due to our definition of intermediate age stars, which spans the range 2 to 10 Gyr. In this particular case, the W09 models recover more of the younger component of the intermediate age populations than sMILES does. Also, note that the youngest W09 model only has an age of 2 Gyr. We speculate that the lack of ages younger than 2 Gyr in W09 might affect the fraction of stars placed in the intermediate age bin. Nevertheless, this requires a dedicated investigation that is beyond the scope of this study. 

\subsection{Total Metallicity}

\begin{figure}
    \centering
    \includegraphics[width=0.8\linewidth]{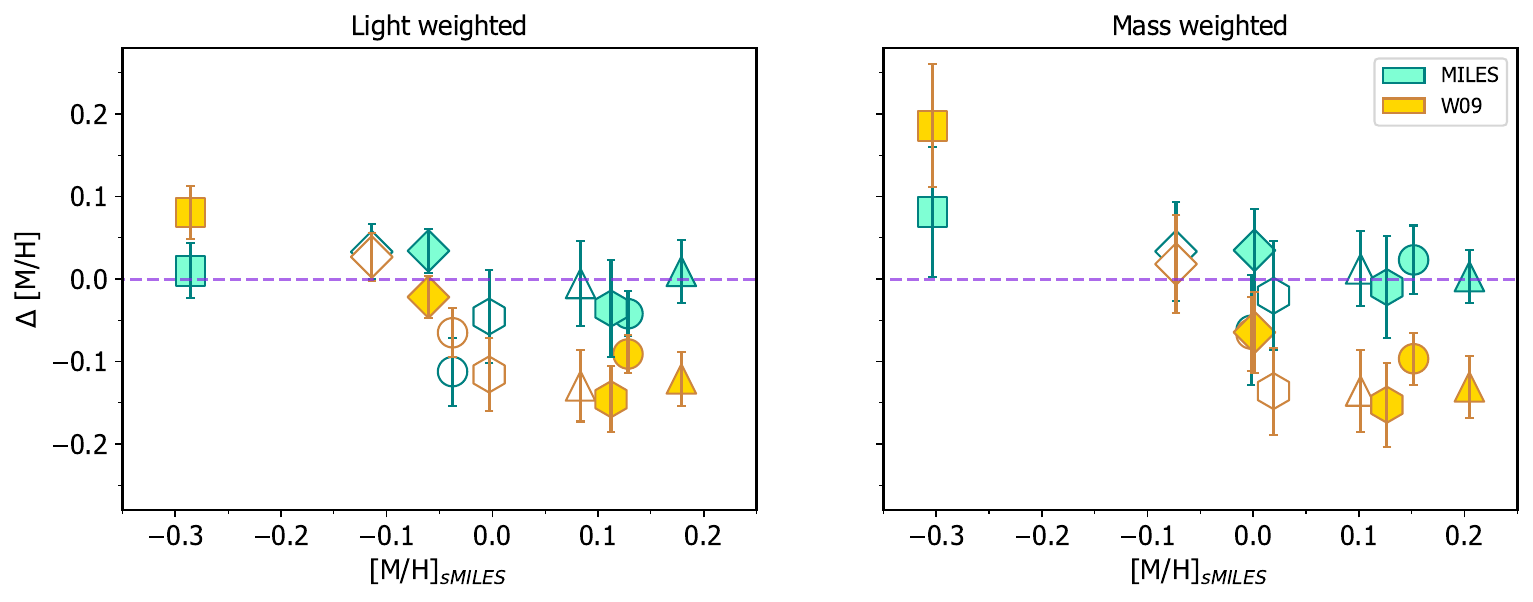}
    \caption{Differences of $[\mathrm{M/H}]$ between the sMILES models and those from the comparison models. Marker descriptions are the same as those in Fig. \ref{fig:age_comparison}.}
    \label{fig:metal_comparison}
\end{figure}

The sMILES metallicities are in a good agreement with the MILES results, as shown in Fig. \ref{fig:metal_comparison}. However, some discrepancies are present compared to the W09 values. The W09 templates give lower metallicity in places where sMILES recover solar metallicity or higher. The exception to this is in the inner region of NGC\,1351, where W09 produces a higher metallicity than sMILES. Whether this is related to the fact that sMILES give a sub-solar metallicity is unclear at this point. The discrepancies also mean that W09 produces a narrower range of total metallicity for our sample, $-0.2 \lesssim [\mathrm{M/H}] \lesssim +0.1$, compared with the range of $-0.3 \lesssim [\mathrm{M/H}] \lesssim +0.2$, recovered by both the MILES and sMILES libraries.  

\subsection{Alpha Abundance}

\begin{figure}
    \centering
    \includegraphics[width=0.8\linewidth]{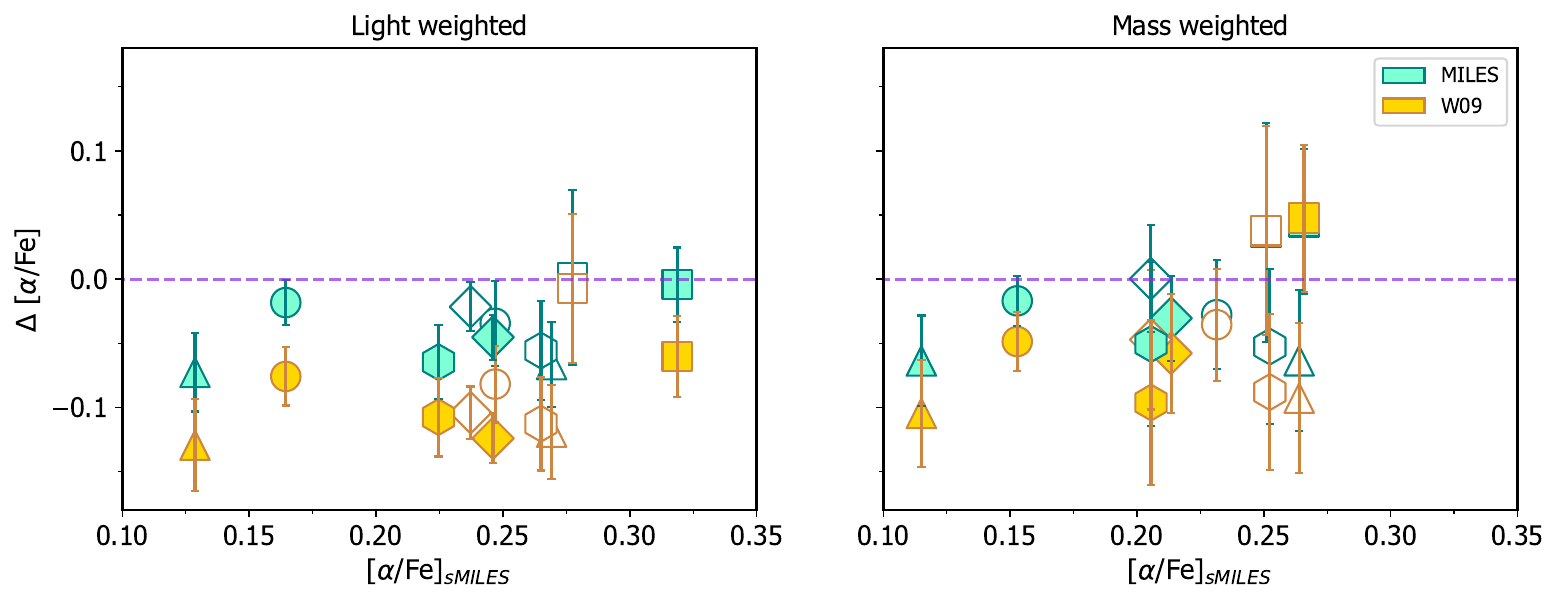}
    \caption{Differences of $[\alpha\mathrm{/Fe}]$ between the sMILES models and those from the comparison models. Marker descriptions are the same as those in Fig. \ref{fig:age_comparison}.}
    \label{fig:alpha_comparison}
\end{figure}

Overall, the sMILES templates produce higher $[\alpha\mathrm{/Fe}]$ abundances than the other models.  This can be seen in Fig. \ref{fig:alpha_comparison}:  the sMILES values 
are systematically higher by $\sim 0.05$ dex than those of MILES and $\sim 0.1$ dex higher than the W09 results. However, since the differences are systematic, the $[\alpha\mathrm{/Fe}]$ trend between the inner and outer region of each galaxy is not affected.

\end{appendix}

\end{document}